\documentclass[12pt,usenames,dvipsnames]{article}

\usepackage{latexsym}
\usepackage{amssymb,amsfonts,amsmath}
\usepackage{graphicx} 
\usepackage{indentfirst}
\usepackage{bbm}
\usepackage{amssymb}
\usepackage{verbatim}
\usepackage{amsmath, amsthm,amssymb}
\usepackage{mathrsfs}
\usepackage{hyperref}
\usepackage{amsfonts}
\usepackage{dsfont}
\usepackage{cite}
\usepackage{xcolor}
\usepackage{enumerate}
\usepackage{cleveref}

\parskip=\medskipamount

\newcommand {\cA}{{\cal A}}

\newcommand {\cD}{{\cal D}}
\newcommand {\cE}{{\cal E}}
\newcommand {\cF}{{\cal F}}

\newcommand {\cJ}{{\cal J}}
\newcommand {\cK}{{\cal K}}
\newcommand {\cL}{{\cal L}}
\newcommand {\cM}{{\cal M}}
\newcommand {\cN}{{\cal N}}

\newcommand {\cR}{{\cal R}}
\newcommand {\cS}{{\cal S}}
\newcommand {\cT}{{\cal T}}
\newcommand {\cU}{{\cal U}}
\newcommand {\cV}{{\cal V}}
\newcommand {\cW}{{\cal W}}

\def\a{\alpha}

\def\b{\beta}

\def\d{\delta}
\def\e{\epsilon}

\def\g{\gamma}

\def\l{\lambda}
\def\m{\mu}

\def\o{\omega}

\def\q{\theta}
\def\r{\rho}
\def\s{\sigma}

\def\x{\xi}
\def\z{\zeta}
\def\D{\Delta}
\def\F{\Phi}

\def\L{\Lambda}
\def\O{\Omega}

\def\S{\Sigma}
\def\U{\Upsilon}

\def\rd{{\rm d}}
\def\ri{{\rm i}}
\def\re{{\rm e}}

\newcommand{\ad}{{\dot{\alpha}}}                           %
\newcommand{\bd}{{\dot{\beta}}}                            %
\newcommand{\ve}{\varepsilon}                            %
\newcommand{\cDB}{{\bar\cD}}                            %

\renewcommand{\aa}{{\a\ad}}
\newcommand{\bb}{{\b\bd}}
\newcommand{\hf}{\frac12}

\newcommand{\vf}{\varphi}

\newcommand{\be}{\begin{equation}}
\newcommand{\ee}{\end{equation}}
\newcommand{\bea}{\begin{eqnarray}}
\newcommand{\eea}{\end{eqnarray}}
\newcommand{\non}{\nonumber}

\newcommand{\bm}[1]{\mbox{\boldmath$#1$}}

\def\double #1{#1{\hbox{\kern-2pt $#1$}}}

\newcommand{\gd}{{\dot\g}}
\newcommand{\dd}{{\dot\d}}

\newcommand{\ts}{{\tilde{\s}}}

\newcommand{\Nabla}{\bm{\nabla}}

\newif\ifdtup

\newcommand{\bsubeq}{\begin{subequations}}
\newcommand{\esubeq}{\end{subequations}}

\numberwithin{equation}{section}

\newcommand{\sSU}{\mathsf{SU}}
\newcommand{\sSL}{\mathsf{SL}}

\newcommand{\sU}{\mathsf{U}}

\begin{document}

\begin{titlepage}
\begin{flushright}
December, 2023 \\
\end{flushright}
\vspace{5mm}

\begin{center}
{\Large \bf 

$\cN=3$ conformal superspace in four dimensions

}
\end{center}

\begin{center}

{\bf Sergei M. Kuzenko and Emmanouil S. N. Raptakis} \\
\vspace{5mm}

\footnotesize{
{\it Department of Physics M013, The University of Western Australia\\
35 Stirling Highway, Perth W.A. 6009, Australia}}  
~\\
\vspace{2mm}
~\\
Email: \texttt{ 
sergei.kuzenko@uwa.edu.au, emmanouil.raptakis@uwa.edu.au}\\
\vspace{2mm}

\end{center}

\begin{abstract}
\baselineskip=14pt
We develop a superspace formulation for ${\cal N}=3$ conformal supergravity in four spacetime dimensions as a gauge theory of the superconformal group $\mathsf{SU}(2,2|3)$. Upon imposing certain covariant constraints, 
the algebra of 
conformally covariant derivatives $\nabla_A = (\nabla_a,\nabla_\a^i,\bar{\nabla}_i^\ad)$ is shown to be determined
in terms of 
a single 
primary chiral spinor superfield, the super-Weyl spinor $W_\a$ of dimension $+1/2$ and its conjugate. 
Associated with $W_\alpha$ is its primary descendant $B^i{}_j$ of dimension $+2$, the super-Bach tensor,
which determines the equation of motion for conformal supergravity.
As an application of this construction, we present two different but equivalent action principles
for ${\cal N}=3$ conformal supergravity. We describe the model for linearised $\cN=3$ conformal supergravity 
in an arbitrary conformally flat background and demonstrate that it possesses $\mathsf{U}(1)$ duality invariance.
Additionally, upon degauging certain local symmetries, our superspace geometry is shown to reduce to the $\mathsf{U}(3)$ superspace constructed by Howe more than four decades ago. Further degauging proves to lead to a new superspace formalism, called $\mathsf{SU}(3) $ superspace, which can also  be used to describe ${\cal N}=3$ conformal supergravity. Our conformal superspace setting opens up the possibility to formulate the dynamics of the off-shell ${\cal N}=3$ super Yang-Mills theory coupled to conformal supergravity. 
\end{abstract}
\vspace{5mm}

\vfill

\vfill
\end{titlepage}

\newpage
\renewcommand{\thefootnote}{\arabic{footnote}}
\setcounter{footnote}{0}

\tableofcontents{}
\vspace{1cm}
\bigskip\hrule

\allowdisplaybreaks

\section{Introduction}

The construction of $\cN=1,2, 4$ conformal supergravity theories in four dimensions 
\cite{KTvN,BdRdW, Butter:2016mtk, Butter:2019edc}
belong to 
the major developments in supergravity,
in particular due to the following:
\begin{itemize}

\item In conjunction with the description of Poincar\'e supergravity as broken superconformal gravity
\cite{KakuTownsend}, the methods of $\cN=1$ and $\cN=2$ conformal supergravity have been used in developing the superconformal tensor calculus to formulate off-shell matter-coupled supergravity
theories in diverse dimensions $d\leq 6$, see \cite{FVP} for a recent review. 

\item 
$\cN=4$ conformal supergravity\footnote{This is the largest conformal supergravity in four dimensions \cite{Ferrara:1977ij, deWit:1978pd}.} 
coupled to four vector multiplets \cite{deRoo} is free of UV divergences and, therefore, anomaly-free. This was originally demonstrated by Fradkin and Tseytlin \cite{FT84,FT85} for the minimal $\cN=4$ conformal supergravity studied in \cite{BdRdW}. The cancellation of anomalies was shown also directly in
\cite{Romer:1985yg}.\footnote{In contrast to the $\cN=4$ case, one cannot cancel divergences/anomalies by coupling $\cN=3$ conformal supergravity to vector multiplets.}
More recently Tseytlin  pointed out \cite{Tseytlin} that analogous conclusions hold for the non-minimal $\cN=4$ conformal supergravities classified in \cite{Butter:2016mtk, Butter:2019edc}.

\item Conformal supergravity gave birth to
conformal higher-spin gauge theory \cite{FT85}.

\end{itemize}

The complete descriptions of $\cN=1$ and $\cN=2$ conformal supergravity theories were originally developed within the component approach \cite{KTvN,BdRdW}, although several superspace formulations appeared in parallel \cite{ZuminoSS, OS, HT, Siegel78}. Remarkably, the story of $\cN=4$ conformal supergravity is quite different. While the  $\cN=4$ Weyl multiplet  was presented more than forty years ago \cite{BdRdW}, the complete actions for $\cN=4$ conformal supergravity (involving a holomorphic function of the complex scalar that parametrises an $\sSU(1, 1) / \sU(1)$ coset space) were constructed only a few years ago \cite{Butter:2016mtk, Butter:2019edc}. One may argue that this was achieved using the $\cN=4$ conformal superspace, which is sketched in the appendices of \cite{Butter:2019edc} and  is a natural generalisation of the $\cN=1$ and $\cN=2$ conformal superspace approaches developed earlier  by Butter \cite{ButterN=1, ButterN=2}.\footnote{These and other formulations for $\cN=1$ and $\cN=2$ conformal supergravity  have recently been reviewed in \cite{Review1,Review2}.}

In general, the superconformal tensor calculus, as reviewed in \cite{FVP}
has proved to be truly useful in formulating general two-derivative supergravity-matter systems 
and studying their dynamical properties. In our opinion, the superconformal setting becomes especially powerful within superspace formulations for supergravity, which: (i) provide remarkably compact expressions for general supergravity-matter actions; (ii) make manifest the geometric properties of such theories; and, most importantly, 
(iii) offer unique tools to generate higher-derivative couplings in matter-coupled supergravity. 
Superspace approaches becomes indispensable  if one has to deal with off-shell 
supermultiplets with infinitely many auxiliary fields, which are: 
(i) the charged $\cN=2$ hypermultiplet \cite{GIKOS, GIOS, LR1, LR2}; and (ii) the $\cN=3$ super Yang-Mills theory \cite{GIKOS1,GIKOS2,RoslyS}.\footnote{Galperin, Ivanov and Ogievetsky \cite{Galperin:1986id} 
derived explicit realisations of the $\cN=3$ superconformal group in the $\cN=3$ harmonic superspace and some of its complex subspaces.}

Recently, the Weyl multiplet for $\cN=3$ conformal supergravity has been constructed by van Muiden and Van Proeyen \cite{vanMvanP}, see also \cite{HS}. Subsequently, the $\cN=3$ conformal supergravity action was derived in the component setting of \cite{HMS} by making use of the superspace methods developed earlier within the framework of $\cN=4 $ conformal supergravity in four dimensions  \cite{Butter:2019edc} as well as the conformal supergravity theories in three \cite{BKNT-M1, BKNT-M2, KNT-M},  five \cite{ BKNT-M15} and six \cite{BKNT} dimensions.
The main goal of the present paper is to develop a superspace counterpart of these results and, more importantly, to provide a framework for coupling the off-shell $\cN=3$ super Yang-Mills multiplet to conformal supergravity. 

The original approach of \cite{BdRdW} 
to describing the $\cN$-extended Weyl multiplet ($\cN \leq 4$) is based on gauging the $\cN$-extended superconformal group in $x$-space. 
Within the conformal superspace approach, developed in \cite{ButterN=1, ButterN=2, Butter:2019edc} for the cases $\cN=1,2,4$ and in the present paper for $\cN=3$,
the gauging of the $\cN$-extended superconformal group is carried out in superspace. 
There is a different superspace approach to $\cN$-extended conformal supergravity, which was developed by Howe practically at the same time as \cite{BdRdW} and is often called the $\sU(\cN)$ superspace \cite{Howe}.\footnote{Its name originates from the fact that its structure group is $\sSL(2,{\mathbb C}) \times \sU(\cN)_R$.} In this superspace setting, the chiral action for $\cN=3$ conformal supergravity was described by M\"uller \cite{Muller}. 
In what follows, we will demonstrate how the $\sU(3)$ superspace formalism is obtained from $\cN=3$ conformal superspace by degauging of certain local symmetries. 

Recently, a new supertwistor formulation was discovered for 
$\cN$-extended conformal supergravity \cite{HL20}.
It is expected to be related to conformal superspace, however relevant technical details have not yet been worked out in the literature.

This paper is organised as follows. In section \ref{Section2}, by gauging the superconformal group $\sSU(2,2|3)$ in superspace and imposing a set of covariant constraints we describe the geometry of $\cN=3$ conformal superspace. As an application of this formalism, in section \ref{Section3} we propose a superconformal action principle based on a chiral Lagrangian superfield. Utilising this principle, we obtain actions for both conformal supergravity and its linearisation about conformally flat backgrounds. Next, in section \ref{Section4}, by making use of on-shell vector multiplet coupled to conformal supergravity, we detail an approach to reduce a given superconformal chiral action to components. The procedure of degauging from conformal superspace to $\sU(3)$, and then to $\sSU(3)$, superspace is described in section \ref{Section5}. Finally, in section \ref{Section6}, we provide some concluding comments and sketch the details of future research directions. The main body of this paper is accompanied two technical appendices. In appendix \ref{AppendixA} we describe our conventions for the $\cN=3$ superconformal algebra $\mathfrak{su}(2,2|3)$. Appendix \ref{AppendixB} is devoted to a study of the $\cN=3$ Weyl multiplet via component reduction.

\section{Conformal superspace}

\label{Section2}

We consider a curved 
superspace $\cM^{4|12}$, parametrised by local coordinates 
$z^{M} = (x^{m},\theta^{\m}_\imath,\bar \theta_{\dot{\mu}}^\imath)$, where $m=0, 1, 2, 3$, $\mu = 1, 2$, $\dot{\mu} = \dot{1}, \dot{2}$ and
$\imath = \underline{1}, \underline{2}, \underline{3}$. 
Its structure group is chosen to be $\sSU(2,2|3)$, the $\cN=3$ superconformal group.
The corresponding Lie superalgebra, $\mathfrak{su}(2,2|3)$, is spanned by the translation $P_A=(P_a, Q_\a^i ,\bar Q^\ad_i)$, Lorentz $M_{ab}$,  $R$-symmetry
$\mathbb{Y}$ and $\mathbb{J}^{i}{}_j$, dilatation $\mathbb{D}$ and the special conformal $K^A=(K^a, S^\a_i ,\bar S_\ad^i)$ generators, see appendix \ref{AppendixA} for more details. 

In order to gauge the superconformal algebra, we associate with each non-translational generator $X_{\underline A} = (M_{ab}, \mathbb{J}^i{}_j, \mathbb{Y}, \mathbb{D}, K^A)$ a connection one-form $\O^{\underline{A}} = (\O^{ab},\F^j{}_i,\F,B,\mathfrak{F}_A) = \rd z^M \O_M{}^{\underline A}$, and with $P_A$ a supervielbein one-form $E^A = \rd z^M E_M{}^A$. We assume that the supermatrix $E_M{}^A$ is non-singular, $\text{Ber}(E_M{}^A) \equiv \text{sdet}(E_M{}^A) \neq 0$, thus a unique inverse exists. The latter is denoted $E_A{}^M$ and satisfies the properties $E_A{}^M E_M{}^B = \d_A{}^B$ and $E_M{}^A E_A{}^N = \d_M{}^N$. With respect to the supervielbein basis, the connection is expressed as $\O^{\underline{A}} = E^B \O_B{}^{\underline A}$, where $\O_B{}^{\underline A} = E_B{}^M \O_M{}^{\underline A}$. 

The conformally covariant derivatives $\nabla_A = (\nabla_a,\nabla_\a^i,\bar{\nabla}_i^\ad)$ are then given by:
\begin{align}
	\label{6.1}
	\nabla_A = E_{A}{}^M \partial_M - \O_A{}^{\underline B} X_{\underline B}~. 
\end{align} 
They obey the graded commutation relations
\begin{align}
	\big [ \nabla_A , \nabla_B \big \} = - \cT_{AB}{}^C \nabla_C -\cR_{AB}{}^{\underline C} X_{\underline C}~,
\end{align}
where $\cT_{AB}{}^C$ is the torsion and $\cR_{AB}{}^{\underline C}$ collectively denotes the curvatures associated with the non-translational generators of $\sSU(2,2|3)$. Additionally, it is assumed that the operators $\nabla_A$ replace $P_A$ and obey the graded commutation relations, c.f. \eqref{StructureConstants},
\begin{align}
[ X_{\underline{B}} , \nabla_A \} = -f_{\underline{B} A}{}^C \nabla_C
- f_{\underline{B} A}{}^{\underline{C}} X_{\underline{C}} ~,
\end{align}

By definition, the gauge group of conformal supergravity is generated by local transformations of the form
\begin{align}
	\label{66.2}
	\nabla_A' = \re^{\mathscr{K}} \nabla_A \re^{-\mathscr{K}} ~, \qquad
	\mathscr{K} =  \xi^B \nabla_B + \L^{\underline B} X_{\underline B} ~.
\end{align}
It follows that the connection one-forms transform via:
\begin{subequations}
	\label{connectionTfs}
	\begin{align}
		\label{connectionTfs-a}
		\delta_{\mathscr K} E^A &= \rd \xi^A + E^B \L^{\underline C} f_{\underline C B}{}^A
		+ \omega^{\underline B} \xi^{C} f_{C \underline B}{}^{A}
		+ E^B \xi^{C} \cT_{C B}{}^A~, \\
		\delta_{\mathscr K} \Omega^{\underline A} &= \rd \L^{\underline A}
		+ \Omega^{\underline B} \L^{\underline C} f_{\underline C \underline B}{}^{\underline A}
		+ \Omega^{\underline B} \xi^{C} f_{C \underline B}{}^{\underline A}
		+ E^B \L^{\underline C} f_{\underline C B}{}^{\underline A}
		+ E^B \xi^{C}  \cR_{C B}{}^{\underline A}~.
	\end{align}
\end{subequations}

Additionally, given a conformally covariant tensor superfield $\cU$ (with its indices suppressed), its corresponding transformation law is
\begin{align}
	\label{6.3}
	\cU' = \re^{\mathscr{K}} \cU~.
\end{align}
Such a superfield is called primary if $K^A \cU =0$. Its dimension $\D$ and $\sU(1)$ charge $Q$ are defined
as follows: ${\mathbb D} \cU = \D \cU$ and ${\mathbb Y} \cU = Q \cU$.

\subsection{Geometric constraints}

In general, the algebra of covariant derivatives $\big [\nabla_A, \nabla_B \big \}$ should be constrained such that it: (i) has a super Yang-Mills structure \cite{Sohnius:1978wk}
\begin{subequations}
	\label{constraints}
\begin{align}
	\label{spinorAlgebra}
	\big \{ \nabla_\a^i , \nabla_\b^j \big \} = 2 \ve_{\a \b} \ve^{i j k} \bar{\mathbb{W}}_k ~, \quad
	\big \{ \nabla_\a^i , \bar{\nabla}_j^\bd \big \} = - 2 \ri \d^i_j \nabla_\a{}^\bd ~, \quad
	\big \{ \bar{\nabla}^\ad_i , \bar{\nabla}^\bd_j \big \} = - 2 \ve^{\ad \bd} \ve_{i j k} \mathbb{W}^k ~,
\end{align}
where $\mathbb{W}^i$ is an operator taking the general form
\begin{align}
	\label{2.2}
	{\mathbb{W}}^i &= {\mathbb{W}}(\nabla)^{i, B} \nabla_B + {\mathbb{W}}(X)^{i, \underline{B}} X_{\underline B} ~,
\end{align}
and its conjugate $\bar{\mathbb{W}}_i$ is defined as $\bar{\mathbb{W}}_i = \overline{\mathbb{W}^i}$;
and (ii) is expressed solely in terms of a single covariant superfield, which is the $\cN=3$ super-Weyl tensor. In accordance with the linearised results \cite{HST}, the latter is expected to be a chiral, primary spinor superfield $W_\a$ of dimension $\hf$ and $\sU(1)_R$ charge $-3$
\begin{align}
	\label{SWeyl}
	\bar{\nabla}^\ad_i W_\a = 0 ~, \qquad K^B W_\a = 0 ~, \qquad \mathbb{D} W_\a = \hf W_\a ~, \qquad \mathbb{Y} W_\a = -3 W_\a~.
\end{align}
\end{subequations}
Below, by studying the consistency conditions associated with these constraints, we elucidate the full geometric structure of conformal superspace.

\subsection{Solution to the Bianchi identities}

We begin by recalling that the covariant derivative $\nabla_A$ satisfies the following Jacobi identities
\begin{align}
	\label{2.3}
	(-1)^{\ve_A \ve_C} \big [\nabla_A , \big [\nabla_B , \nabla_C \big \} \big \} + (\text{two cycles}) = 0~,
\end{align}
where $\ve_A$ denotes the parity of $\nabla_A$. In the absence of the geometric constraints imposed in the previous subsection, condition \eqref{2.3} leads to a set of equations, known as the Bianchi identities, which are identically satisfied. However, upon imposing the constraints \eqref{constraints}, this is no longer the case and instead they must be solved to elucidate the superspace's geometric structure. The present subsection is devoted to such an analysis.

At dimension-$\frac{3}{2}$, the Jacobi identity \eqref{2.3} leads to the constraints
\begin{subequations}
	\label{2.4}
\begin{align}
	\label{2.4a}
	\big [\nabla_\a^{(i} , {\mathbb{W}}^{j)} \big] = 0 ~, \qquad \big[\bar{\nabla}^\ad_{i} , {\mathbb{W}}^{j} \big] = \frac{1}{3} \d_i^j \big [\bar{\nabla}^\ad_{k} , {\mathbb{W}}^{k} \big] ~,
\end{align}
and determines the following commutation relations
\begin{align}
	\label{2.4b}
	\big [ \nabla_\a^i , \nabla_{\bb} \big ] = - \frac{\ri}{2} \ve_{\a \b} \ve^{i j k} \big [ \bar{\nabla}_{\bd j}, \bar{\mathbb W}_k \big ] ~, \qquad \big [ \bar{\nabla}^\ad_i , \nabla_{\bb} \big ] = - \frac{\ri}{2} \d^\ad_\bd \ve_{i j k} \big [ \nabla_\b^j , \mathbb{W}^k \big ] ~.
\end{align}
Next, at dimension-$2$, we obtain the `reality' condition 
\begin{align}
	\label{2.4c}
	 \ve_{jkl} \big\{ \nabla^{\a i} , \big [ \nabla_\a^k , \mathbb{W}^l \big] \big \} = \ve^{ikl} \big\{ \bar{\nabla}_{\ad j} , \big [\bar{\nabla}^{\ad}_k, \bar{\mathbb W}_l \big] \big\} ~,
\end{align}
and the commutation relation for vector covariant derivatives
\begin{align}
	\label{2.4d}
	\big [ \nabla_{\aa} , \nabla_{\bb} \big ] = \frac{1}{12} \ve_{\ad \bd} \ve_{i j k} \big \{ \nabla_{(\a}^i \big [ \nabla_{\b)}^j , \mathbb{W}^k \big ]\big \} + \frac{1}{12} \ve_{\a \b} \ve^{ijk} \big \{ \bar{\nabla}_{(\ad i} \big [ \bar{\nabla}_{\bd) j} , \bar{\mathbb{W}}_k \big ]\big \} ~.
\end{align}
\end{subequations}
It should be emphasised that the above relations encode complete information regarding the Jacobi identities \eqref{2.3}; given equations \eqref{2.4}, one may show that all remaining implications of \eqref{2.3} immediately follow. Further, the operator $\mathbb{W}^i$ and its conjugate $\bar{\mathbb{W}}_i$ completely determine the algebra of covariant derivatives for this supergeometry.

Additionally, by studying the following Jacobi identity
\begin{align}
	\label{2.5}
	(-1)^{\ve_{\underline{A}} \ve_C} \big [X_{\underline{A}} , \big [\nabla_B , \nabla_C \big \} \big \} + (\text{two cycles}) = 0~,
\end{align}
one may show that $\mathbb{W}^i$ is a primary, dimension-$1$ operator of $\sU(1)_R$ charge $-2$
\begin{align}
	\label{2.6}
	\big [K^B , \mathbb{W}^i \big] = 0 ~, \qquad \big [\mathbb{D} , \mathbb{W}^i \big ] = \mathbb{W}^i ~, \qquad [\mathbb{Y} , \mathbb{W}^i \big ] = - 2 \mathbb{W}^i ~.
\end{align}
Thus, on dimensional grounds, it is clear that $\mathbb{W}(\nabla)^{i,B}$ must take the form
\begin{align}
	\label{2.8}
	\mathbb{W}(\nabla)^{i,b} = 0 ~, \qquad \mathbb{W}(\nabla)^{i,}{}_{j}^{\b} = \d^{i}_j W^\b ~, \qquad \mathbb{W}(\nabla)^{i,}{}_{\bd}^j = 0 ~,
\end{align}
where $W_\a$ is the super-Weyl tensor, see equation \eqref{SWeyl}.

Next, we will show that, by studying \eqref{2.8} in tandem with equations \eqref{2.4} and \eqref{2.6}, all components of $\mathbb{W}^i$ defined in \eqref{2.2} may be solved for in terms of $W_\a$. The solution is:
\begin{subequations}
	\begin{align}
		\mathbb{W}(M)^{i,ab} &= (\s^{ab})_{\a \b} \mathbb{W}(M)^{i,\a \b} ~, \qquad \mathbb{W}(M)^{i,\a \b} = \nabla^{(\a i } W^{\b)}~, \\
		\mathbb{W}(\mathbb{J})^{i,j}{}_k &= \d^i_k \nabla^{\a j} W_\a - \frac 1 3 \d^j_k \nabla^{\a i} W_\a ~, \\
		\mathbb{W}(\mathbb{Y})^i &= \frac 1 2 \nabla^{\a i} W_\a ~, \qquad
		\mathbb{W}(\mathbb{D})^i = - \frac {1}{12} \nabla^{\a i} W_\a~, \\
		\mathbb{W}(S)^{i,}{}_{\b}^j &= \frac 1 4 \nabla_{\b}^{i} \nabla^{\g j} W_\g~, \qquad \mathbb{W}(\bar{S})^{i,}{}_{j}^{\bd} = \frac{\ri}{2} \d^i_j \nabla^{\bb} W_\b~, \\
		\mathbb{W}(K)^{i,}{}_{\bb} &= - \frac 1 2 \nabla^\a{}_\bd \nabla^{i}_\b W_\a~.
	\end{align}
\end{subequations}
As a result, the algebra of covariant derivatives $[\nabla_A, \nabla_B\}$ is now completely determined in terms of $W_\a$. For completeness, we present these relations below (up to dimension-$\frac 3 2$)
\begin{subequations} \label{algebra}
\begin{align}
	\big \{ \nabla_\a^i , \nabla_\b^j \big \} &= 2 \ve_{\a \b} \ve^{i j k} \bigg \{ \bar{W}_\ad \bar{\nabla}^\ad_k - \bar{\nabla}^\ad_k \bar{W}^{\bd} \bar{M}_{\ad \bd} + \bar{\nabla}^\ad_l \bar{W}_\ad \mathbb{J}^l{}_k - \frac{1}{4} \bar{\nabla}^\ad_k \bar{W}_\ad \Big(2\mathbb{D} + \frac13 \mathbb{Y}\Big ) \non \\
	& \qquad \qquad \qquad - \frac \ri 2 \nabla_\g{}^\ad \bar{W}_\ad S^\g_k - \frac 1 4 \bar{\nabla}^\ad_k \bar{\nabla}_l^\bd \bar{W}_\bd \bar{S}_\ad^l + \frac 1 2 \nabla^{\g \bd} \bar{\nabla}^\gd_k \bar{W}_\bd K_{\g \gd} \bigg \} ~, \\	%
	\big \{ \nabla_\a^i , \bar{\nabla}_j^\bd \big \} &= - 2 \ri \d^i_j \nabla_\a{}^\bd ~, \\	%
	\big \{ \bar{\nabla}^\ad_i , \bar{\nabla}^\bd_j \big \} &= - 2 \ve^{\ad \bd} \ve_{i j k} \bigg \{ {W}^\a {\nabla}_\a^k + {\nabla}^{\a k} {W}^{\b} {M}_{\a \b} + {\nabla}^{\a l} {W}_\a \mathbb{J}^k{}_l + \frac{1}{4} {\nabla}^{\a k} {W}_\a \Big(2\mathbb{D} - \frac13 \mathbb{Y}\Big ) \non \\
	& \qquad \qquad \qquad + \frac 1 4 {\nabla}^{k}_\a {\nabla}^{\b l} {W}_\b {S}^\a_{l} + \frac \ri 2 \nabla^{\a \gd} {W}_\a \bar{S}_\gd^k + \frac 1 2 \nabla^{\b \gd} {\nabla}^{\g k} {W}_\b K_{\g \gd} \bigg \} ~, \label{algebra-c}\\	%
	\big [ \nabla_\a^i , \nabla_{\bb} ] &= - \ri \ve_{\a \b} \ve^{i j k} \bigg \{ \Big( \bar{\nabla}_{(\bd j} \bar{W}_{\gd)} + \hf \ve_{\bd \gd} \bar{\nabla}^\dd_j \bar{W}_\dd \Big) \bar{\nabla}^\gd_k - \hf \Big( \bar{\nabla}_{\bd j} \bar{\nabla}^\gd_k \bar{W}^\dd + \d_\bd^\gd \bar{\nabla}^\dd_j \bar{\nabla}^{\dot \epsilon}_k \bar{W}_{\dot \epsilon} \Big) \bar{M}_{\gd \dd} \non \\
	& \qquad \qquad \qquad + \bar{\nabla}_{\bd j} \bar{\nabla}^\gd_l \bar{W}_\gd \mathbb{J}^l{}_k - \frac{1}{4} \bar{\nabla}_{\bd j} \bar{\nabla}^\gd_k \bar{W}_\gd \Big(2\mathbb{D} + \frac{1}{3} \mathbb{Y} \Big) - \frac{1}{8} \bar{\nabla}_{\bd j} \bar{\nabla}^\gd_k \bar{\nabla}^\dd_l \bar{W}_\dd \bar{S}^l_\gd \non \\
	& \qquad \qquad \qquad - \frac{\ri}{4} \Big( \bar{\nabla}_{\bd j} \nabla_{\gamma \gd} - 2 \nabla_{\gamma \gd} \bar{\nabla}_{\bd j} \Big) \bar{W}^\gd S^\g_k + \frac 1 4 \bar{\nabla}_{\bd j} \nabla^{\g \dd} \bar{\nabla}^\gd_k \bar{W}_\dd K_{\g \gd}\bigg \} \non \\
	& \quad ~+ 2 \ri \ve_{\a \b}  \bar{W}_\bd \bigg \{ {W}^\g {\nabla}_\g^i + {\nabla}^{\g i} {W}^{\d} {M}_{\g \d} + {\nabla}^{\g j} {W}_\g \mathbb{J}^i{}_j + \frac{1}{4} {\nabla}^{\g i} {W}_\g \Big(2\mathbb{D} - \frac13 \mathbb{Y}\Big ) \non \\
	& \qquad \qquad \qquad + \frac 1 4 {\nabla}^{k}_\g {\nabla}^{\d l} {W}_\d {S}^\g_{l} + \frac \ri 2 \nabla^{\g \gd} {W}_\g \bar{S}_\gd^k + \frac 1 2 \nabla^{\g \dd} {\nabla}^{\d i} {W}_\g K_{\d \dd} \bigg \} ~,\\
	\big [ \bar{\nabla}^\ad_i , \nabla_{\bb} \big ] &= \ri \d^\ad_\bd \ve_{i j k} \bigg \{ \Big( \nabla_{(\b}^j W_{\g)} - \hf \ve_{\b \g} \nabla^{\d j} {W}_\d \Big) {\nabla}^{\g k} - \hf \Big( \nabla_\b^j \nabla^{\g k} W^\d + \d_\b^\g \nabla^{\d j} \nabla^{\e k} W_\e \Big) {M}_{\g \d} \non \\
	& \qquad \qquad \qquad - {\nabla}_{\b}^j \nabla^{\g l} W_\g \mathbb{J}^k{}_l - \frac{1}{4} \nabla_\b^j \nabla^{\g k} W_\g \Big(2 \mathbb{D} - \frac 1 3 \mathbb{Y}\Big) - \frac{1}{8} \nabla_\b^j \nabla_\g^k \nabla^{\d l} W_\d S^\g_l \non \\
	& \qquad \qquad \qquad - \frac{\ri}{4} \Big({\nabla}_{\b}^j \nabla^{\g \gd} - 2 \nabla^{\g \gd} \nabla_\b^j \Big)W_\g \bar{S}_\gd^k + \frac 1 4 \nabla_\b^j \nabla^{\d \gd} \nabla^{\g k} W_\d K_{\g \gd}\bigg \} \non \\
	& \quad - 2 \ri \d^{\ad}_{\bd}  {W}_\b \bigg \{ \bar{W}_\gd \bar{\nabla}^\gd_i - \bar{\nabla}^\gd_i \bar{W}^{\dd} \bar{M}_{\gd \dd} + \bar{\nabla}^\gd_j \bar{W}_\gd \mathbb{J}^j{}_i - \frac{1}{4} \bar{\nabla}^\gd_i \bar{W}_\gd \Big(2\mathbb{D} + \frac13 \mathbb{Y}\Big ) \non \\
	& \qquad \qquad \qquad - \frac \ri 2 \nabla_\g{}^\gd \bar{W}_\gd S^\g_k - \frac 1 4 \bar{\nabla}^\gd_i \bar{\nabla}_j^\dd \bar{W}_\dd \bar{S}_\gd^j + \frac 1 2 \nabla^{\d \gd} \bar{\nabla}^\dd_i \bar{W}_\gd K_{\d \dd} \bigg \}  ~.
\end{align}
\end{subequations}
We emphasise that the commutator of vector derivatives, $[\nabla_\aa , \nabla_\bb]$, is completely determined by these relations, in conjunction with equation \eqref{2.4d}.

The structure of this algebra leads to highly non-trivial implications. In particular, given a primary chiral superfield, $\bar{\nabla}^\ad_i \F = 0$, consistency of \eqref{algebra-c} with the superconformal algebra relation \eqref{A.5d} implies that: (i) it carries no dotted spinor indices $(\bar{M}_{\ad \bd} \F = 0$); and, (ii) its dimension $\D$ and $\sU(1)$ charge $Q$ are related by $Q = -6 \D$.

At this stage, all Bianchi identities have been solved, with the sole exception of the eq. \eqref{2.4c}. A routine analysis shows it yields the following constraint
\begin{align}
	\ri \ve_{jkl} \nabla^{ \a i} \nabla_\a^k \nabla^{\b l} W_\b - \frac{\ri}{3} \d^i_j \ve_{klm} \nabla^{ \a k} \nabla_\a^l \nabla^{\b m} W_\a  = \ri \ve^{ikl} \bar{\nabla}_{\ad j} \bar{\nabla}_k^\ad  \bar{\nabla}_{\bd l} \bar{{W}}^\bd - \frac \ri 3 \d^i_j \ve^{klm} \bar{\nabla}_{\ad k} \bar{\nabla}_l^\ad  \bar{\nabla}_{\bd m} \bar{W}^\bd  ~.
\end{align}
This is equivalent to the requirement that the superfield ${B} = ({B}^i{}_j)$, defined by
\begin{align}
	\label{superBach}
	{B}^i{}_j :=
	\ri \ve_{jkl} \nabla^{ \a i} \nabla_\a^k \nabla^{\b l} W_\b - \frac{\ri}{3} \d^i_j \ve_{klm} \nabla^{ \a k} \nabla_\a^l \nabla^{\b m} W_\a  ~,
\end{align}
is Hermitian and traceless
\begin{subequations}
	\label{2.15}
\begin{align}
	{B}^{\dagger} = {B} ~, \qquad \text{tr} \, {B} = 0 ~,\label{2.14b}
\end{align}
with the superconformal properties
\begin{align}
	K^C {B}^i{}_j = 0 ~, \qquad \mathbb{D} {B}^i{}_j = 2 {B}^i{}_j ~, \qquad \mathbb{Y} {B}^i{}_j = 0~, \label{2.14c}
\end{align}
and satisfies the conservation equations
\begin{align}
	\label{2.14d}
	\nabla_\a^{(i} {B}^{j)}{}_k = \frac{1}{4} \d_k^{(i} \nabla_\a^{|l|} {B}^{j)}{}_l ~, \qquad \bar{\nabla}_{(i}^\ad {B}^{j}{}_{k)} = \frac{1}{4} \d_{(i}^j \bar{\nabla}_{|l|}^{\ad} {B}^l{}_{k)}
\end{align}
\end{subequations}

We note that the equations \eqref{2.15} imply that ${B}^{i}{}_j$ has the properties of  a conformal $\cN=3$ supercurrent ${\mathfrak J}^i{}_j$.\footnote{The constraints on  ${\mathfrak J}^i{}_j$ were described in \cite{HST} and \cite{HKR}
in flat and conformally flat backgrounds, respectively. 
In conformal superspace, the properties of ${\mathfrak J}^i{}_j$ are given by eqs.
\eqref{2.15} with $B^i{}_j$  replaced with ${\mathfrak J}^i{}_j$.
}
In particular, since ${B}^i{}_j$  is a descendant of the super-Weyl tensor $W_\a$, it is the
$\cN=3$ supersymmetric extension of the Bach tensor.

\section{Action functionals and conformal supergravity}

\label{Section3}

Having introduced a manifestly superconformal geometric approach to $\cN=3$ conformal supergravity above, the goal of this section will be the construction of a superspace action for this theory. In order to do so, we will first give the salient details of a superconformal action principle based on a chiral Lagrangian function below, before reviewing linearised conformal supergravity and subsequently extending our analysis to the full nonlinear theory.

\subsection{Chiral action principle}
\label{Section3.1}
To construct an off-shell model for conformal supergravity, a superconformal action principle is required. While actions can be defined as integrals over the full superspace or its chiral subspace, the former approach proves to be difficult to work with in the present case as the Lagrangian must carry negative conformal dimension. It is for this reason that we will restrict our attention only to actions defined over the chiral subspace.

Given a primary chiral scalar Lagrangian $\cL_c$ of dimension $+1$,
\begin{subequations}\label{3.2}
\begin{align}
	K^A \cL_c = 0 ~, \qquad \bar{\nabla}^\ad_i \cL_c = 0 ~, \qquad \mathbb{D} \cL_c = \cL_c ~,
\end{align}
it may be shown that the chiral action
\begin{align}
	\cS = \int \rd^{4}x \, \rd^{6} \q \, \cE \, \cL_c ~,
\end{align}
\end{subequations}
is invariant under general gauge transformations \eqref{6.3}, hence it is locally superconformal. Here $\cE$ denotes the chiral superspace measure. Below, we will make use of this action principle to postulate an action for conformal supergravity and its linearisation about a conformally flat supergeometry.

\subsection{Linearised conformal supergravity}

We begin by studying conformal supergravity linearised about a conformally flat background. Such a superspace is characterised by the vanishing of the super-Weyl tensor; $W_\a = 0$. Linearised conformal supergravity is then described by a `reduced' chiral field strength $\mathfrak{W}_\a$ 
\begin{subequations}
	\label{LinSW}
\begin{align}
	\label{LinSW-a}
	\bar{\nabla}^\ad_i \mathfrak{W}_\a = 0~, \qquad \ri \ve_{jkl} \nabla^{ik} \nabla^{\a l} \mathfrak{W}_\a = \ri \ve^{ikl} \bar{\nabla}_{jk}  \bar{\nabla}_{\ad l} \bar{\mathfrak{W}}^\ad ~,
\end{align}
where we have used the notational shorthand
\begin{align}
	\nabla^{ij} = \nabla^{\a (i} \nabla_{\a}^{j)} ~, \qquad \bar{\nabla}_{ij} = \bar{\nabla}_{\ad (i} \bar{\nabla}_{j)}^\ad ~.
\end{align}
The superconformal transformation law of \eqref{LinSW-a} is characterised by the properties (cf. \eqref{SWeyl}):
\begin{align}
	K^B \mathfrak{W}_\a = 0~, \qquad \mathbb{D} \mathfrak{W}_\a = \hf \mathfrak{W}_\a ~, \qquad \mathbb{Y} \mathfrak{W}_\a = - 3 \mathfrak{W}_\a~.
\end{align}
\end{subequations}
These constraints allow one to express the linearised super-Weyl tensor $\mathfrak{W}_\a$ in terms of an unconstrained gauge prepotential $H^i{}_j$ as follows
\begin{align}
	\mathfrak{W}_{\a} = \ve_{ikl}\bar{\nabla}^6 \nabla^{jk} \nabla_\a^l H^{i}{}_j~,
\end{align}
where the chiral operator $\bar{\nabla}^6$ takes the form
\begin{align}
	\bar{\nabla}^6 = - \frac{1}{1152} \ve^{ijk} \ve^{lmn} \bar{\nabla}^{\ad_1}_i \bar{\nabla}^{\ad_2}_j \bar{\nabla}^{\ad_3}_k \bar{\nabla}_{(\ad_1 l} \bar{\nabla}_{\ad_2 m} \bar{\nabla}_{\ad_3) n} = - \frac{1}{1152} \ve^{ijk} \ve^{lmn} \bar{\nabla}_{il} \bar{\nabla}_{jm} \bar{\nabla}_{kn} ~.
\end{align}
The superfield $H^{i}{}_j$ is Hermitian, traceless and defined modulo the gauge transformations
\begin{align}
	\d_\z H^i{}_j = \nabla_\a^i \z^\a_j + \bar{\nabla}^\ad_j \bar{\z}_\ad^i - \frac 1 3 \d_j^i \Big( \nabla_\a^k \z^\a_k + \bar{\nabla}^\ad_k \bar{\z}_\ad^k \Big)~,
\end{align}
which are consistent with superconformal symmetry provided it satisfies
\begin{align}
	K^C H^i{}_j = 0 ~, \qquad \mathbb{D} H^i{}_j = -2 H^{i}{}_j ~, \qquad \mathbb{Y} H^{i}{}_j = 0~.
\end{align}

From $\mathfrak{W}_\a$, and its conjugate $\bar{\mathfrak{W}}_\ad$, one can construct the superconformal gauge-invariant functional
\begin{align}
	\label{LinCSG}
	\cS_\text{LCSG}[\mathfrak{W},\bar{\mathfrak{W}}] = \hf \int \rd^{4}x \, \rd^{6} \q \, \cE \, \mathfrak{W}^\a \mathfrak{W}_{\a} + \text{c.c.} ~,
\end{align}
which describes propagation of the linearised conformal supergravity multiplet in conformally flat backgrounds. Its overall numerical coefficient has been chosen in accordance with the identity
\begin{align}
	\label{3.7}
	\ri \int \rd^{4}x \, \rd^{6} \q \, \cE \, \mathfrak{W}^\a \mathfrak{W}_{\a} + \text{c.c.} = 0~,
\end{align}
which holds up to a total derivative. Our action functional \eqref{LinCSG} extends the one proposed by Siegel in 1981 \cite{Siegel} to arbitrary conformally flat backgrounds.

It may be shown that the model \eqref{LinCSG} possesses $\sU(1)$ duality invariance. This is most easily seen by noting that the equation of motion for $H^i{}_j$ is
\begin{align}
	\label{3.8}
	\ri \ve_{jkl} \nabla^{ik} \nabla^{\a l} \mathfrak{M}_\a = \ri \ve^{ikl} \bar{\nabla}_{jk}  \bar{\nabla}_{\ad l} \bar{\mathfrak{M}}^\ad ~, 
\end{align}
where $\mathfrak{M}_\a = - \ri \mathfrak{W}_\a$.
Hence, the off-shell constraint \eqref{LinSW-a} and dynamical equation above are together invariant under the rigid $\sU(1)$ rotations 
\begin{align}
	\label{U(1)rot}
	\d_\l \mathfrak{W}_\a = \l \mathfrak{M}_\a ~, \qquad \d_\l \mathfrak{M}_\a = - \l \mathfrak{M}_\a ~, \qquad \l \in \mathbb{R}~.
\end{align}
This realisation serves as a starting point in constructing nonlinear extensions of linearised conformal supergravity. 

Such a duality-invariant nonlinear theory is described by an action $\cS_{\text {NL}}[\mathfrak{W},\bar{\mathfrak{W}}]$, which is assumed to depend only on $\mathfrak{W}_\a$ and its conjugate. Considering this as a functional of the chiral, but otherwise unconstrained primary superfield $\mathfrak{W}_\a$ (and its conjugate $\bar{\mathfrak{W}}_\ad$), we may define the dual chiral field strength
\begin{align}
	\label{3.10}
	\ri \mathfrak{M}_{\a} := \frac{\d \cS_{\text {NL}}[\mathfrak{W},\bar{\mathfrak{W}}]}{\d \mathfrak{W}^\a} ~, \qquad \bar{\nabla}^\ad_i \mathfrak{M}_{\a} = 0~, \qquad K^B \mathfrak{M}_\a = 0 ~, \qquad \mathbb{D} \mathfrak{M}_\a = \hf \mathfrak{M}_\a~,
\end{align}
where the variational derivative is defined as follows
\begin{align}
	\d \cS_{\text {NL}}[\mathfrak{W},\bar{\mathfrak{W}}] = \int \rd^{4}x \, \rd^{6} \q \, \cE \, \d \mathfrak{W}^\a \frac{\d \cS_{\text {NL}}[\mathfrak{W},\bar{\mathfrak{W}}]}{\d \mathfrak{W}^\a} + \text{c.c.}
\end{align}
Now, varying $\cS_{\text {NL}}[\mathfrak{W},\bar{\mathfrak{W}}]$ with respect to $H^i{}_j$ yields the dynamical equations \eqref{3.8}, where $\mathfrak{M}_\a$ is defined in eq. \eqref{3.10}. This model may then be shown to possess $\sU(1)$ duality invariance \eqref{U(1)rot} provided the following self-duality equation holds
\begin{align}
	\text{Im} \int \rd^{4}x \, \rd^{6} \q \, \cE \, \Big \{ \mathfrak{W}^\a \mathfrak{W}_{\a} + \mathfrak{M}^\a \mathfrak{M}_{\a} \Big \} =0~,
\end{align}
where $\mathfrak{W}_\a$ is taken to be a general chiral spinor.

Recently, the general formalism of duality rotations for $\cN$-extended superconformal gauge multiplets was described in \cite{Kuzenko:2023ebe}, see also \cite{Kuzenko:2021qcx}. In the $\cN=3$ case, such models were described in terms of chiral field strengths $\mathfrak{W}_{\a(m+n+3)}$, $m,n\geq0$. We emphasise that here we have described duality-invariant models for a chiral spinor field strength $\mathfrak{W}_\a$, which was not considered in \cite{Kuzenko:2023ebe}.

\subsection{Conformal supergravity}

Building on the construction of linearised conformal supergravity above, in this subsection we deduce the superspace action for $\cN=3$ conformal supergravity. We begin by noting that the latter must: (i) be locally superconformal; (ii) be built in terms of the covariant superfield $W^\a$; and (iii) reduce to the linearised model \eqref{LinCSG} in the appropriate limit. Hence, the action functional for $\cN=3$ conformal supergravity must take the form
\begin{align}
	\label{3.9}
	\cS_{\text{CSG}}[W,\bar{W}] = \hf \int \rd^{4}x \, \rd^{6} \q \, \cE \, W^\a W_{\a} + \text{c.c.}
\end{align}
It may be shown that the dynamical equation resulting from this action is simply
\begin{align}
	B^i{}_j = 0~,
\end{align}
i.e. that the geometry is super-Bach flat. Thus, it follows that every conformally flat supergeometry is a solution to conformal supergravity.
Additionally, upon degauging to the $\sU(3)$ superspace described in section \ref{section5.1}, our action \eqref{3.9} reduces to the one given by M\"uller \cite{Muller}.

Further, as a consequence of equation \eqref{3.7}, it is expected that the functional
\begin{align}
	\label{3.11}
	\mathfrak{P} = \frac{\ri}{2} \int \rd^{4}x \, \rd^{6} \q \, \cE \, W^\a W_{\a} + \text{c.c.} ~,
\end{align}
is a topological invariant. Specifically, it should define an $\cN=3$ supersymmetric generalisation of the Pontryagin topological invariant. A proof of this claim will be given elsewhere.

\section{Alternative action principle}
\label{Section4}

We now describe an approach to reduce a given superconformal chiral action to components. This procedure makes use of the 
the superform (or ectoplasm) approach 
to the construction of supersymmetric invariants 
 \cite{Hasler, Gates, GGKS}.\footnote{The approach proves equivalent 
to the rheonomic formalism \cite{Castellani}.} 
 It has been applied within the conformal superspace framework in diverse dimensions
 \cite{BKNT-M2, KNT-M, BKNT-M15, BKNT}. Here we give the salient details of the method.

\subsection{Primary closed four-forms in superspace}

Central to this method in four spacetime dimensions is 
a super four-form 
\begin{subequations}
\label{4.1}
\begin{align}
	\cJ = \frac{1}{4!} E^{A_4} \wedge E^{A_3} \wedge E^{A_2} \wedge E^{A_1} \cJ_{A_1 A_2 A_3 A_4} ~, \qquad E^A = \rd z^M E_M{}^A~,
\end{align}
that is constrained to be both primary,\footnote{Some care is needed when studying \eqref{4.1Primary} as the supervielbein one-forms $E^A$ transform non-covariantly under the action of the special conformal generators, see \cite{BKNT-M2,BKNT} for similar discussions in three and six dimensions.
These transformations follow from \eqref{connectionTfs-a}.} 
\begin{align}
	\label{4.1Primary}
	K^A \cJ = 0~,
\end{align}
and closed,
\begin{align}
	\rd \cJ = 0 \quad \iff \quad \nabla_{[B} \cJ_{A_1 A_2 A_3 A_4 \}} + 2 \cT_{[B A_1}{}^{C} \cJ_{|C| A_2 A_3 A_4 \} } = 0~.
	\label{4.1b}
\end{align}
\end{subequations}
These conditions ensure that the following functional
\begin{align}
	\label{4.2}
	\cS = \frac{1}{4!} \int \rd^{4}x \, \ve^{m_1 m_2 m_3 m_4} E_{m_4}{}^{A_4} E_{m_3}{}^{A_3} E_{m_2}{}^{A_2} E_{m_1}{}^{A_1} \cJ_{A_1 A_2 A_3 A_4} ~,
\end{align}
has the properties: (i) it is independent of the Grassmann coordinates $(\cS = \cS|_{\q = \bar{\q} = 0})$; and (ii) it is locally superconformal. In the case that $\cJ$ is not exact 
(that is, it cannot be represented as $\cJ = \rd \U$, where the three-form $\U$ is a well defined local operator of the dynamical variables), then \eqref{4.2} defines a superconformal action. 

There are techniques to construct closed superforms in diverse dimensions, see e.g.
\cite{Arias:2014ona, Gates:2014cqa, Linch:2014iza} and references therein. 
Below we will apply the four-dimensional construction proposed in \cite{GKT-M} to derive closed four-forms in the $\cN=1$ and $\cN=2$ cases. This construction makes use of an {\it on-shell} vector multiplet with $\cF= \hf E^B \wedge E^A \cF_{AB}$ the corresponding gauge-invariant {\it primary} field strength. Since $\rd \cF =0$, 
the primary four-form $\cF \wedge \cF$ obeys \eqref{4.1b}. Additionally, as the vector multiplet is on-shell, $\cF$ can be written as the sum of self-dual and anti-self-dual parts, $\cF = \cF_+ + \cF_-$, which are  related to each other only by complex conjugation (no differential constraint), and therefore $\rd \cF_+ =0$ and $\rd \cF_- =0$. In the $\cN=1$ case,  $\cF_+ $ is expressed in terms of the chiral spinor field strength $\cW_\a$, while $\cF_-$ is expressed in terms of its conjugate $\bar \cW_\ad$. In the $\cN=2$ case,  $\cF_+ $ is expressed in terms of the chiral scalar field strength $\cW$, while $\cF_-$
is expressed in terms of its conjugate $\bar \cW$. Now, $\cF \wedge \cF$ can be written as a sum of three parts, $\cF \wedge \cF =\cF_+ \wedge \cF_+ 
+ \cF_- \wedge \cF_- +2 \cF_+ \wedge \cF_-$, each of which is annihilated by the operator $\rd$. It turns out that $\cF_+ \wedge \cF_+ $ can be expressed in terms of a single primary chiral scalar $\cL_c$ which is $\cW^\a \cW_\a$ in the $\cN=1$ case and $\cW^2$ for $\cN=2$. Viewed as a function of $\cL_c$, the four-form $\cF_+ \wedge \cF_+ $ is  closed and primary, and it cannot be written as the exterior derivative of a three-form.\footnote{The primary four-form $\cF_+ \wedge \cF_- $ is expressed in terms of a conformal supercurrent, which is $\cW_\a \bar \cW_\ad$ in the $\cN=1$ case and $\cW \bar \cW$ for $\cN=2$. } 
A similar construction proves to work in $\cN=3$ conformal superspace.

\subsection{On-shell vector multiplet in conformal supergravity}
\label{Section4.2}

In the present case, the closed superform's construction will be sketched by making use of an on-shell vector multiplet \cite{Sohnius:1978wk} coupled to conformal supergravity, which is described by a real primary and closed two-form
\begin{subequations}
\begin{align}
	\cF = \hf E^B \wedge E^A \cF_{AB} ~, \qquad K^A \cF = 0~, \qquad \rd \cF = 0 ~,
\end{align}
satisfying the following constraints
\begin{align}
	\cF_{\a}^i{}_{\b}^j = - 2 \ve_{\a \b} \ve^{i j k} \bar{\cW}_k ~, \qquad \cF_\a^i{}_j^\bd = 0~, \qquad \cF^{\ad}_i{}^{\bd}_{j} = 2 \ve^{\ad \bd} \ve_{i j k} \cW^k~.
\end{align}
\end{subequations}
Here the complex isospinor $\cW^i$ is a primary superfield of dimension $1$ and $\sU(1)_R$ charge $-2$
\begin{subequations}
	\label{4.4}
\begin{align}
	K^B \cW^i = 0 ~, \qquad \mathbb{D} \cW^i = \cW^i~, \qquad \mathbb{Y} \cW^i = -2 \cW^i ~,
\end{align}
and is subject to the differential constraints
\begin{align}
	\label{4.4b}
	\nabla_\a^{(i} \cW^{j)} = 0 ~, \qquad \bar{\nabla}^\ad_i \cW^j = \frac 1 3 \d_i^j \bar{\nabla}^\ad_k \cW^k,
\end{align}
\end{subequations}
which imply that $\cW^i$ is on-shell. Employing the closure condition $\rd \cF = 0$, a routine calculation allows one to show that the remaining non-zero components of $\cF$ are:
\begin{subequations}
\begin{align}
	\cF_{a}{}^j_\b =& - \frac \ri 4 (\s_a)_{\bb} \Big( \ve^{jkl} \bar{\nabla}^\bd_k \bar{\cW}_l - 4 \bar{W}^\bd \cW^j \Big) ~, \qquad
	\cF_{a}{}^\bd_j = \frac \ri 4 \Big( (\tilde{\s}_a)^{\bd \b} \ve_{jkl} \nabla_\b^k \cW^l + 4 W_\b \bar{\cW}_j \Big) ~, \\
	\cF_{a b} =& - \frac{1}{24}(\s_{ab})^{\a \b} \Big ( \ve_{ijk} \nabla_\a^i \nabla_\b^j \cW^k - 4 W_{\a} \nabla_{\b}^i \bar{\cW}_i  + 12 \nabla_{\a}^i W_{\b} \bar{\cW}_i \Big) \non \\
				&+ \frac 1 {24} (\tilde{\s}_{ab})^{\ad \bd} \Big ( 12 \bar{\nabla}_{\ad i} \bar{W}_\bd \cW^i - 4 \bar{W}_\ad \bar{\nabla}_{\bd i} \cW^i - \ve^{ijk} \bar{\nabla}_{\ad i} \bar{\nabla}_{\bd j} \bar{\cW}_k \Big )~.
\end{align}
\end{subequations}
Thus, $\cF$ takes the form $\cF = \cF_+ + \cF_-$, where
\begin{align}
	\label{4.6}
	\cF_+ =& \bar{E}_\bd^j \wedge \bar{E}_\ad^i \Big( 2 \ve^{\ad \bd} \ve_{ijk} \cW^k\Big) + \bar{E}_\bd^j \wedge E^a \Big( \frac \ri 4 (\tilde{\s}_a)^\bb \ve_{jkl} \nabla_\b^k \cW^l \Big) + E^\b_j \wedge E^a \Big( \ri (\s_a)_\bb \bar{W}^\bd \cW^j\Big) \non \\
	& + E^b \wedge E^a \Big \{ - \frac 1 {24} (\s_{ab})^{\a \b} \ve_{ijk} \nabla_\a^i \nabla_\b^j \cW^k + \hf (\tilde{\s}_{ab})^{\ad \bd} \Big( \bar{\nabla}_{\ad i} \bar{W}_\bd \cW^i - \frac 1 3 \bar{W}_\ad \bar{\nabla}_{\bd i} \cW^i \Big ) \Big \} ~.
\end{align}

Now, by inspecting eq. \eqref{4.4b}, we see that the two superfields $\cW^i$ and $\bar{\cW}_i$ are independent at the level of the equations of motion. Further, it is clear from eq. \eqref{4.6} and so the two sectors involving $\cW^i$ and its conjugate are independent, modulo the reality condition.
By construction, $\rd \cF =0$, which implies $\rd \cF_+ =0$.

\subsection{The main construction}
\label{section4.3}

By making use of the two-form \eqref{4.6}, we now turn to analysing the closed four-form $\cJ = \cF_+ \wedge \cF_+$. 
Direct calculations lead to
\begin{align}
	\cJ =&~ \bar{E}^l_\dd \wedge \bar{E}^k_\gd \wedge \bar{E}^j_\bd \wedge \bar{E}^i_\ad \Big( 4 \ri \ve^{\ad \bd} \ve^{\gd \dd} \ve_{ijm} \ve_{kln} \mathscr{L}^{mn} \Big) \non \\
	&+ \bar{E}^l_\dd \wedge \bar{E}^k_\gd \wedge \bar{E}^j_\bd \wedge E^a \Big( \frac{2}{3}  \ve^{\gd \dd} \ve_{jnp} \ve_{klm} (\tilde{\s}_a)^\bb \nabla_\b^n \mathscr{L}^{mp} \Big) \non \\
	&+ \bar{E}^l_\dd \wedge \bar{E}^k_\gd \wedge E^\b_j \wedge E^a \Big( -4 (\s_a)_\bb \ve_{klm} \bar{W}^\bd \mathscr{L}^{jm} \Big) \non \\
	&+ \bar{E}^l_\dd \wedge \bar{E}^k_\gd \wedge E^b \wedge E^a \Big( \frac \ri{16} (\s_{ab})^{\a \b} \ve^{\gd \dd} \ve_{ijm} \ve_{kln} \nabla_\a^i \nabla_\b^j \mathscr{L}^{mn} + \frac \ri {12} (\tilde{\s}_{ab})^{\gd \dd} \ve_{kmn} \ve_{lpq} \nabla^{mi} \mathscr{L}^{nj} \non \\
	& \qquad \qquad \qquad \qquad \qquad \qquad - 2 \ri (\tilde{\s}_{ab})^{\ad \bd} \ve^{\gd \dd} \ve_{klm} \Big[ \bar{\nabla}_{\ad n} \bar{W}_\bd - \frac 1 4 \bar{W}_\ad \bar{\nabla}_{\bd n} \Big ]\mathscr{L}^{mn} \Big)\non \\
	&+ E^\d_l \wedge {E}^\g_k \wedge E^b \wedge E^a \Big( - \ri (\s_{ab})_{\g \d}  \bar{W}^2 \mathscr{L}^{kl}  \Big) \non \\
	&+ \bar{E}_\dd^l \wedge {E}^\g_k \wedge E^b \wedge E^a \Big( \frac{\ri}{3} (\s_a)_{\g \gd} (\tilde{\s}_b)^{\d \dd} \ve_{lmn} \bar{W}^\gd \nabla_\d^m \mathscr{L}^{kn}  \Big) \non \\
	&+ \bar{E}_\dd^l \wedge {E}^c \wedge E^b \wedge E^a \ve_{abcd} \Big( \ri (\s^d)^{\a \dd} \Big [ - \frac 1 {192} \ve_{ijk} \ve_{lmn} \nabla^{\b m} \nabla_{(\a}^i \nabla_{\b)}^j \mathscr{L}^{kn} + \frac{5}{1152} \ve_{lij} \bar{\nabla}^\ad_k \bar{W}_\ad \nabla_\a^i \mathscr{L}^{jk} \Big]  \Big) \non \\
	&+ {E}^\d_l \wedge {E}^c \wedge E^b \wedge E^a \ve_{abcd} 
	\Big( \frac 1{48} 
	(\s^d)^\a{}_\bd \ve_{ijk} \bar{W}^\dd \nabla_\a^i \nabla_\d^j \mathscr{L}^{kl} - \frac 1 3 
	 (\s^d)_\d{}^\ad \bar{W}^\bd \bar{\nabla}_{(\ad i} \bar{W}_{\bd)} \mathscr{L}^{il}  \non \\
	& \qquad \qquad \qquad \qquad \qquad \qquad - \frac 1 {16} 
	(\s^d)_\d{}^\ad \bar{W}^2 \bar{\nabla}_{\ad i} \mathscr{L}^{il} \Big) \non \\
	&+ {E}^d \wedge {E}^c \wedge E^b \wedge E^a 
	\ve_{abcd}
	\Big(- \frac{1}{4608} \ve_{ijk} \ve_{lmn} \nabla^{\a l} \nabla^{\b m} \nabla_{(\a}^i \nabla_{\b)}^j \mathscr{L}^{kn} - \frac{5}{864} \bar{\nabla}^\ad_i \bar{W}_\ad \bar{\nabla}^\bd_j \bar{W}_\bd \mathscr{L}^{ij} \non \\
	& \qquad \qquad \qquad \qquad \qquad \qquad + \frac{5}{3456} \bar{\nabla}^\ad_i \bar{W}_\ad \bar{W}_\bd \bar{\nabla}^\bd_j \mathscr{L}^{ij} + \frac{1}{24} \bar{\nabla}^{(\ad}_i \bar{W}^{\bd)} \bar{\nabla}_{\ad j} \bar{W}_\bd \mathscr{L}^{ij} \Big) \non \\
	& \qquad \qquad \qquad \qquad \qquad \qquad - \frac{1}{48} \bar{\nabla}_{(\ad i} \bar{W}_{\bd)} \bar{W}^\ad \bar{\nabla}^\bd_j \mathscr{L}^{ij} - \frac{1}{384} \bar{W}^2 \bar{\nabla}_{ij} \mathscr{L}^{ij} \Big) ~.
\end{align}
Here we have made the definition $\mathscr{L}^{ij} := - \ri \cW^{i} \cW^j$. In accordance with equations \eqref{4.4}, this superfield is characterised by the following superconformal properties
\begin{subequations}
\label{4.8}
\begin{align}
	K^C \mathscr{L}^{ij} = 0 ~,\qquad \mathbb{D} \mathscr{L}^{ij} = 2 \mathscr{L}^{ij} ~, \qquad  \mathbb{Y} \mathscr{L}^{ij} = -4 \mathscr{L}^{ij}~,
\end{align}
and differential constraints
\begin{align}
	\nabla_\a^{(i} \mathscr{L}^{jk)} = 0~, \qquad \bar{\nabla}^\ad_i \mathscr{L}^{jk} = \frac{1}{2} \d_i^{(j} \bar{\nabla}^\ad_l \mathscr{L}^{k) l} ~.
\end{align}
\end{subequations}
Remarkably, these features allow us to relate  $\mathscr{L}^{ij}$ to a general chiral Lagrangian $\cL_c$, eq. \eqref{3.2}, by the rule:
\bea
\mathscr{L}^{ij} = \nabla^{ij} \cL_c~. 
\eea

At the heart of this action principle is the multiplet $\mathscr{L}^{ij}$, thus it is instructive to study its $\cN=2$ superfield content. Such an analysis requires us to switch off the super-Weyl spinor, $W_\a = 0$. Let $\Nabla_\a^{\hat i}$ and $\bar{\Nabla}^\ad_{\hat i}$ be the spinor covariant derivatives of $\cN=2$ conformal superspace where $\hat{i} = \underline{1}, \underline{2}$ is an $\sSU(2)_R$ index. They may be defined in terms of their $\cN=3$ counterparts as follows: $\Nabla_\a^{\hat i} {\mathfrak U} =  \nabla_\a^{\hat{i}} U|$
and $\bar{\Nabla}^\ad_{\hat i} {\mathfrak U}= \bar{\nabla}^{\ad}_{\hat{i}} U|$.
Here $U$ is a covariant $\cN=3$ superfield, and  ${\mathfrak U} \equiv U| :=U|_{\theta^\a_{\underline 3} = \bar{\theta}_\ad^{\underline 3} = 0}$ is its $\cN=2$ projection.
By a routine analysis, we see that $\mathscr{L}^{ij}$ contains the following 
 independent $\cN=2$ superfields in its multiplet
\begin{subequations}
	\begin{align}
		\mathfrak{G}^{\hat{i} \hat{j}} &= 
		\mathscr{L}^{\hat{i} \hat{j}} 
		 | ~, 
		\qquad  \Nabla_\a^{(\hat i} \mathfrak{G}^{\hat{j} \hat{k})} = 0 ~, \quad \bar{\Nabla}_\ad^{(\hat{i}} \mathfrak{G}^{\hat{j} \hat{k})} = 0~,\\
		\mathfrak{J}^{\hat i} &= \mathscr{L}^{\hat{i} \underline{3}} | ~, \qquad \Nabla^{(\hat{i} \hat{j}} \mathfrak{J}^{\hat k)} = 0~, \qquad \bar{\Nabla}_\ad^{(\hat i} \mathfrak{J}^{\hat j)} = 0~,\\
		\mathfrak{L} &= \mathscr{L}^{\underline{3} \underline{3}} | ~, \qquad \bar{\Nabla}_\ad^{\hat i} \mathfrak{L} = 0~.
	\end{align}
\end{subequations}
We see that $\mathscr{L}^{ij}$  encodes: (i) a complex linear multiplet $\mathfrak{G}^{\hat{i} \hat{j} } $;
(ii) a supercurrent $\mathfrak{J}^{\hat i}$ associated with the $\cN=2$ superconformal gravitino multiplet \cite{HKR}; and (iii) a chiral scalar $\mathfrak{L}$ which defines a chiral Lagrangian.

\section{Degauging to conventional superspace}
\label{Section5}

According to eq. \eqref{66.2}, under an infinitesimal special superconformal gauge transformation $\mathscr{K} = \Lambda_{B} K^{B}$, the dilatation connection transforms as follows
\bea
\d_{\mathscr{K}} B_{A} = - 2 \Lambda_{A} ~.
\eea
Thus, it is possible to impose the gauge
$B_{A} = 0$, which completely fixes 
the special superconformal gauge freedom.
As a result, the corresponding connection is no longer required for the covariance of $\nabla_A$ under the residual gauge freedom and
should be separated,
\bea
\nabla_{A} &=& \cD_{A} - \mathfrak{F}_{AB} K^{B} ~. \label{ND}
\eea
Here the operator $\cD_{A} $ involves only the Lorentz and $R$-symmetry connections
\bea
\cD_A = E_A{}^M \partial_M - \frac{1}{2} \O_A{}^{bc} M_{bc} - \Phi_A{}^j{}_k \mathbb{J}^{k}{}_j - \ri \F_A \mathbb{Y}~.
\eea

While the $\sSU(2,2|3)$ structure group symmetry has been partially broken by our imposition of the gauge $B_A=0$ above, there is still some residual symmetry present. In particular, the transformations we are left with with include local $\cK$-transformations of the form
\begin{align}
	\nabla_A' = \re^{\mathcal{K}} \nabla_A \re^{-\mathcal{K}} ~, \qquad
	\mathcal{K} =  \xi^B \cD_B+ \hf K^{bc} M_{bc} + \ri \rho \mathbb{Y} 
	+ \chi^{i}{}_j \mathbb{J}^{j}{}_i ~ = \bar{\mathcal{K}} ~,
\end{align}
which act on tensor superfields (with indices suppressed) as $\cU' = \re^{\mathcal{K}} \cU$. Thus, we see that the local structure group has broken to $\sSL(2,\mathbb{C}) \times \sU(3)_R$. It should also be mentioned that the above transformations are not the most general conformal supergravity gauge transformations preserving the gauge $B_A=0$. Specifically, there is a class of residual combined dilatation and special conformal transformations preserving this constraint. As will be shown in the following subsection, these generate super-Weyl transformations.

Now, by making use of the following relation
\bea
\label{4.3}
\big [ \cD_{A} , \cD_{B} \big \} &=&  \big [ \nabla_{A} , \nabla_{B} \big \} + \big(\cD_{A} \mathfrak{F}_{BC} - (-1)^{\ve_A \ve_B} \cD_{B} \mathfrak{F}_{AC} \big) K^C + \mathfrak{F}_{AC} \big [ K^{C} , \nabla_B \big \} \non \\
&& - (-1)^{\ve_A \ve_B} \mathfrak{F}_{BC} \big [ K^{C} , \nabla_A \big \} - (-1)^{\ve_B \ve_C} \mathfrak{F}_{AC} \mathfrak{F}_{BD} \big [K^D , K^C \big \} ~,
\eea
it is possible to describe the torsion and curvature tensors associated with $\cD_A$ in terms of $W_\a$ and the `degauged' special conformal connection $\mathfrak{F}_{AB}$. Such an analysis is performed below.

\subsection{$\sU(3)$ superspace geometry}\label{section5.1}

Employing equation \eqref{4.3}, a routine calculation leads to the following expressions for the degauged special conformal connection:
\begin{subequations} \label{connections}
	\bea
	\mathfrak{F}_\a^i{}_\b^j  
	&=&
	-\hf\ve_{\a\b}S^{ij}
	- \ve^{ijk }Y_{\a\b k}
	+ \frac 1 4 \ve_{\a \b} \ve^{ijk} \bar{\cD}^\ad_k \bar{W}_\ad
	~,
	\\
	\mathfrak{F}^\ad_i{}^\bd_j  
	&=&
	-\hf\ve^{\ad\bd}\bar{S}_{ij}
	+ \ve_{ijk} \bar{Y}^{\ad\bd k}
	- \frac 14 \ve^{\ad \bd} \ve_{ijk} \cD^{\a k} W_\a
	~,\\
	\mathfrak{F}_\a^i{}^\bd_j
	&=&
	- \mathfrak{F}^\bd_j{}_\a^i
	=
	-\d^i_jG_\a{}^\bd
	-\ri G_\a{}^\bd{}^i{}_j
	~,
	\\
	\mathfrak{F}_{\a}^{i}{}_{,\bb}
	&=&  \ri \cD_{(\a}^i G_{\b) \bd} - \frac{1}{4} \cD_{(\a}^j G_{\b) \bd}{}^i{}_j + \ve_{\a \b} \Big ( \frac{\ri}{4} \ve^{ijk} \bar{\cD}_{(\bd j} \bar{\cD}_{\gd) k} \bar{W}^\gd - \ri \bar{Y}_{\bd \gd}^i \bar{W}^\gd - \frac \ri 4 \cD^{\g i} W_\g \bar{W}_\bd \non \\
	&~&+ \frac \ri 2 \cD^{\g i} G_{\g \bd} + \frac 1 4 \cD^{\b j} G_{\b \gd}{}^{i}{}_j \Big ) ~,
	\\
	\mathfrak{F}_{\bb}{}_{,\a}^i &=& \mathfrak{F}_{\a}^{i}{}_{,\bb} + \ri \ve_{\a \b} \Big ( \frac{1}{4} \ve^{ijk} \cDB_{\bd j} \cDB_k^\gd \bar{W}_\gd + \cD^{\g i} W_\g \bar{W}_\bd + 2 \bar{Y}_{\bd \gd}^i \bar{W}^\gd \Big )~, ~~~
	\\
	\mathfrak{F}_{\ad i ,\bb}
	&=&
	- \ri \bar{\cD}_{(\ad i} G_{\b \bd)} + \frac{1}{4} \bar{\cD}_{(\ad j} G_{\b) \bd}{}^j{}_i + \ve_{\ad \bd} \Big ( \frac \ri 4 \ve_{ijk} \cD^j_{(\b} \cD_{\g)}^k W^\g + \ri Y_{\b \g i} W^\g + \frac{\ri}{4} \bar{\cD}^\gd_i \bar{W}_\gd W_{\b} \non \\
	&~&- \frac \ri 2 \bar{\cD}^\gd_{ i} G_{\b \gd} - \frac{1}{4} \bar{\cD}_{j}^\gd G_{\b \gd}{}^j{}_i \Big )~,
	\\
	\mathfrak{F}_{\bb, \ad i} &=& \mathfrak{F}_{\ad i ,\bb} +  \ri\ve_{\ad \bd} \Big( \frac{1}{4} \ve_{ijk} \cD_\b^j \cD_\g^k W^\g + W_{\b} \cDB^\gd_i \bar{W}_\gd - 2 Y_{\b \g i} W^\g \Big)
	~, ~~~~~~~~~~ \\
	\mathfrak{F}_{\aa,\bb} &=& \frac{\ri}{6} \Big( 
	\cD_\a^i \mathfrak{F}_{\ad i,\bb} + \bar{\cD}_{\ad i} \mathfrak{F}^i_{\a,\bb} + 2 \ri \Big( \mathfrak{F}_\a^i{}_\b^j \mathfrak{F}_{\ad i , \bd j}
	+ \mathfrak{F}^i_{\a, \bd j} \mathfrak{F}_{\ad i, \b}{}^j
	\Big)
	\Big)~.
	\eea
\end{subequations}
The dimension-1 superfields introduced above have the following symmetry properties:  
\bea
S^{ij}=S^{ji}~, \qquad Y_{\a\b i} =Y_{\b\a i}~, \qquad {G_{\a \ad}{}^{i}{}_i} = 0~,
\eea
and satisfy the reality conditions
\bea
\overline{S^{ij}} =  \bar{S}_{ij}~,\qquad
\overline{Y_{\a\b i}} = \bar{Y}_{\ad\bd}^i~,\qquad
\overline{G_{\b\ad}} = G_{\a\bd}~,\qquad
\overline{G_{\b\ad}{}^{i}{}_j} = - G_{\a\bd}{}^j{}_{i}
~.~~~~~~
\eea
The ${\sU}(1)_R$ charges of the complex fields are:
\bea
{\mathbb Y} S^{ij}=2S^{ij}~,\qquad
{\mathbb Y}  Y_{\a\b i}=2Y_{\a\b i}~.
\eea
Now, by employing \eqref{4.3}, we find that the anti-commutation relations for the spinor covariant derivatives are:
\begin{subequations} \label{U(2)algebra}
	\bea
	\{ \cD_\a^i , \cD_\b^j \}
	&=&
	2 \ve_{\a \b} \ve^{ijk} \bar{W}_\ad \bar{\cD}^\ad_k
	+ 4 S^{ij}  M_{\a\b} 
	- 4 \ve_{\a \b} \ve^{ijk} Y^{\g \d}_k M_{\g \d}
	+ 2 \ve_{\a \b} \ve^{ijk} \bar{\cD}^\ad_k \bar{W}^\bd \bar{M}_{\ad \bd}
	\non \\
	&~&
	- 4\ve_{\a \b} S^{k[i} \mathbb{J}^{j]}{}_k
	+ 8 \ve^{kl(i}{Y}_{\a\b k}  \mathbb{J}^{j)}{}_l
	+ \ve_{\a \b} \ve^{ijk} \bar{\cD}^\ad_l \bar{W}_\ad \mathbb{J}^l{}_k
	~,
	\label{U(2)algebra.a}
	\\
	\{ \cD_\a^i , \bar{\cD}^\bd_j \}
	&=&
	- 2 \ri \d_j^i\cD_\a{}^\bd
	+4\Big(
	\d^i_jG^{\g\bd}
	+\ri G^{\g\bd}{}^i{}_j
	\Big) 
	M_{\a\g} 
	+4\Big(
	\d^i_jG_{\a\gd}
	+\ri G_{\a\gd}{}^i{}_j
	\Big)  
	\bar{M}^{\bd\gd}
	\non\\
	&&
	+8 G_\a{}^\bd \mathbb{J}^i{}_j
	+4\ri\d^i_j G_\a{}^\bd{}^{k}{}_j \mathbb{J}^i{}_{k}
	-\frac{2}{3}\Big(
	\d^i_jG_\a{}^\bd
	+\ri G_\a{}^\bd{}^i{}_j
	\Big)
	\mathbb{Y} 
	~,
	\label{U(2)algebra.b} \\
	\{ \cDB_i^\ad , \cDB_j^\bd \}
	&=&
	- 2 \ve^{\ad \bd} \ve_{ijk} W^\a \cD_\a^k
	- 4 \bar{S}_{ij}  \bar{M}^{\ad\bd} 
	+ 4 \ve^{\ad \bd} \ve_{ijk} \bar{Y}^{\gd \dd k} \bar{M}_{\gd \dd}
	+ 2 \ve^{\ad \bd} \ve_{ijk} {\cD}^{\a k} {W}^\b {M}_{\a \b}
	\non \\
	&~&
	+ 4\ve^{\ad \bd} \bar{S}_{k[i} \mathbb{J}^{k}{}_{j]}
	- 8 \ve_{kl(i} \bar{Y}^{\ad\bd k}  \mathbb{J}^{l}{}_{j)}
	- \ve^{\ad \bd} \ve_{ijk} \cD^{\a l} {W}_\ad \mathbb{J}^k{}_j
	~,
	\label{U(2)algebra.c}
	\eea
	\esubeq
	At the same time, the consistency conditions arising from solving \eqref{4.3} lead to the Bianchi identities:
	\begin{subequations}\label{BI-U2}
		\bea
		\cD_{\a}^{(i}S^{jk)}&=&0~, \\
		\bar{\cD}_{\ad i}S^{jk} &=& \ri\cD^{\b (j}G_{\b\ad}{}^{k)}{}_i + \frac{1}{4} \d_i^{(j} \Big( 2 \bar{\cD}_{\ad l} S^{k)l} - \ri {\cD}^{\b |l|} G_{\b \ad}{}^{k)}{}_l\Big) - \d^{(i}_j \ve^{k)lm} \cDB_{(\ad l} \cDB_{\bd) m} \bar{W}^\bd ~, ~~~
		\\
		\cD_{(\a}^{i}Y_{\b\g)j}&=& \frac 1 3 \d^i_j \cD_{(\a}^{k} Y_{\b\g)k}, \\
		\cD_\a^k S^{ij} &=& \cD_\a^{(i} S^{j)k} + 2 \ve^{kl(i} \Big ( \cD^{\b j)} Y_{\a \b l} + 3 \ri G_{\aa}^{j)}{}_l \bar{W}^\ad \Big )~, \\
		\cD^{\b i} Y_{\a \b i} &=& \frac{3 \ri}{2} (\cD_\aa - 6\ri G_{\aa}) \bar{W}^\ad \\
		\cD_{(\a}^i G_{\b)\bd} &=& \frac{1}{4} \ve^{ijk} \cDB_{\bd j} Y_{\a \b k} - \frac{\ri}{16} \cD_{(\a}^i G_{\b) \bd}{}^i{}_j ~, \\
		\cD^{\a i} G_{\a \ad}&=& \frac{5\ri}{16} \cD^{\a j} G_{\aa}{}^i{}_j + \frac 1 8 \cDB_{\ad j} S^{ij} - \frac 1 8 \ve^{ijk} \cDB_{\ad j} \cDB_{\bd k} \bar{W}^\bd + \bar{Y}_{\ad \bd}^i \bar{W}^\bd - \frac 3 4 \cD^{\a i} W_\a \bar{W}_\ad ~,~~~ \\
		\cD_{(\a}^{(i}G_{\b)\bd}{}^{j)}{}_k&=&\frac{1}{4} \cD_{(\a}^l G_{\b) \bd}{}^{(i}{}_l \d^{j)}_k~, \\
		\cD_{(\a}^{[i} G_{\b) \bd}{}^{j]}{}_{k} &=& - \frac{1}{2} \cD_{(\a}^{l} G_{\b) \bd}{}^{[i}{}_{l} \d_k^{j]} + \d^{[i}_j \ve^{k]lm} \cDB_{\bd l} Y_{\a\b m}\\
		\cD^{\a[i}G_{\a \ad}{}^{j]}{}_k&=&-\frac{1}{2} \cD^{\a l} G_{\a \ad}{}^{[i}{}_l \d^{j]}_k + \frac{\ri}{4} \ve^{ijl} \Big( \bar{\cD}_{lk} + 2 \bar{S}_{lk} \Big ) \bar{W}_\ad~, ~~~~~ 
		\eea
	\end{subequations}
	where we have made the definitions:
	\begin{align}
		\cD^{ij} = \cD^{\a (i} \cD_{\a}^{j)} ~, \qquad \bar{\cD}_{i j} = \bar{\cD}_{\ad (i}\bar{\cD}^{\ad}_{j)}~.
	\end{align}
	Additionally, the super-Bach tensor \eqref{superBach} takes the form
	\begin{align}
		{B}^i{}_j =& \ve_{jkl} \cD^{\a i} \cD_\a^k \cD^{\b l} W_\b - 4 Y_{\a \b j} \cD^{\a i} W^\b + 6 \ve_{jkl} S^{ik} \cD^{\a l} W_\a
		- \cD^{\a i} W_\a \cDB_{\ad j} \bar{W}^\ad  \non \\
		& - 8 \cD^{\a i} Y_{\a \b j} W^{\b} - 8 G_{\aa}{}^{i}{}_j \bar{W}^\a \bar{W}^\ad - \frac 13 \Big ( \ve_{klm} \cD^{\a k} \cD_{\a}^l \cD^{\b m} W_{\b} - 4 Y_{\a \b k} \cD^{\a k} W^\b \non \\
		& - \cD^{\a k} W_\a \cD_{\ad k} \bar{W}^\ad - 8 \cD^{\a k} Y_{\a \b k} W^{\b} \Big ) ~.
	\end{align}

	Now, we describe how the residual dilation symmetries of conformal superspace manifest in $\sU(3)$ superspace as super-Weyl transformations. It may be shown that the following combined dilatation and special conformal transformation, see eq. \eqref{66.2}, parametrised by a dimensionless real scalar superfield $\S$ = $\bar{\S}$, preserves the gauge $B_A = 0$:
	\begin{align}
		\mathscr{K}(\S) = \S \mathbb{D} + \hf \nabla_B \S K^{B} \quad \implies \quad B'_A = 0~.
	\end{align}
	At the level of the degauged geometry, this induces the following super-Weyl transformations
	\begin{subequations}
		\label{6.17}
		\bea
		\cD^{' i}_\a&=&\re^{\hf \S} \Big( \cD_\a^i+2\cD^{\b i}\S M_{\a \b} + 2 \cD_{\a}^j \S \mathbb{J}^{j}{}_i 
		- \frac{1}{6} \cD_\a^i\S {\mathbb Y} \Big)
		~,
		\label{Finite_D}
		\\
		\bar{\cD}_{i}^{' \ad}&=&\re^{\hf \S} \Big( \bar{\cD}^\ad_i+2 \bar{\cD}_{i}^{\bd} \S \bar{M}_{\ad \bd} - 2 \bar{\cD}^{\ad}_j \S \mathbb{J}^{j}{}_i 
		+ \frac{1}{6} \bar{\cD}_{i}^{\ad} {\mathbb Y} \Big)
		~,
		\label{Finite_DB}\\
		{\cD}'_\aa&=&\re^{\S} \Big( \cD_{\a \ad} + {\rm i} \cD^i_{\a} \S \cDB_{\ad i} + {\rm i} \cDB_{\ad i} \S \cD_{\a}^i + \Big( \cD^\b{}_\ad \S - \ri \cD^{\b i} \S \cDB_{\ad i} \S \Big) M_{\a \b} \non \\ && + \Big( \cD_{\a}{}^\bd \S + \ri \cD_{\a}^i \S \cDB^{\bd}_i \S \Big) { \bar M}_{\ad \bd} 
		- 2\ri \cD_\a^i \S \cDB_{\ad j} \S \mathbb{J}^j{}_i
		+ \frac{\ri}{6} \cD_{\a}^i \S \cDB_{\ad i} \S \mathbb{Y} 
		\Big)
		~,
		\label{Finite_DBB}
		\\
		W^{'}_\a&=& \re^{\hf \S} W_\a
		\\
		S^{' ij}&=& \re^{\S} \Big( S^{ij}
		-\hf \cD^{ij} \S + \cD^{\a (i} \S \cD_\a^{j)} \S \Big)
		\label{Finite_S}~,
		\\
		Y^{'}_{\a\b i}&=& \re^{\S} \Big( Y_{\a\b i}
		+ \frac 14 \ve_{ijk} \cD_\a^{j} \cD_\b^{k} \S + \hf \ve_{ijk} \cD_\a^{j} \S \cD_\b^{k} \S \Big)
		\label{Finite_Y}~,
		\\
		G'_{\a\ad}&=&
		\re^{\S} \Big( G_{\a\ad}
		-{\frac{1}{12}}[\cD_\a^i,\bar{\cD}_{\ad i}]\S
		-\frac{1}{3} \cD_\a^i \S \bar{\cD}_{\ad i} \S \Big)
		~,
		\label{Finite_G}
		\\
		G'_{\a\ad}{}^{i}{}_j&=& \re^{\S} \Big( G_{\a\ad}{}^{i}{}_j
		+{\frac\ri 4} \Big( [\cD_\a^{i},\bar{\cD}_{\ad j}] - \frac{1}{3} \d^i_j [\cD_\a^{k},\bar{\cD}_{\ad k}] \Big) \S \Big)
		~.
		\label{Finite_Gij}
		\eea
	\end{subequations}
	
	It should be mentioned that, in the infinitesimal limit $\S^2 = 0$, the corresponding super-Weyl transformations coincide with the ones of \cite{Howe}. Additionally, in the conformally flat limit, these transformations reduce to the ones given in \cite{KKR}.

	\subsection{Coupling to conformal compensator}
	
	The torsion superfield $G_{\aa}{}^i{}_j$ introduced above turns out to describe purely gauge degrees of freedom. To illustrate this point, we begin by coupling the background supergeometry to some nowhere-vanishing scalar superfield $\Xi \neq 0$ of non-vanishing dimension and $\sU(1)_R$ charge which will play the role of a conformal compensator. By performing an appropriate super-Weyl transformation
	\begin{align}
		\label{CComp}
		\Xi \longrightarrow \re^{\D_\Xi \S + \ri q_{\Xi} \r} \Xi ~, \qquad \D_\Xi, q_\Xi \neq 0~,
	\end{align}
	it is possible to impose the gauge
	\begin{align}
		\Xi = 1~.
	\end{align}
	As one would expect, associated with this gauge condition are several non-trivial integrability conditions. In particular, we see that $G_{\aa}{}^i{}_j$ vanishes in this gauge
	\begin{align}
		\Big ( \big \{ \cD_\a^i , \bar{\cD}_{\ad j} \big \} - \frac 1 3 \d^i_j \big \{ \cD_\a^k , \bar{\cD}_{\ad k} \big \} \Big) \Xi = - \frac{2 q_\Xi}{3} G_\aa{}^{i}{}_j = 0 \quad \implies \quad G_{\aa}{}^i{}_j = 0~.
	\end{align}

	Since $G_{\aa}{}^i{}_j$ is inert under $\sU(1)_R$ transformations, this condition may be obtained simply by performing an appropriate super-Weyl transformation \eqref{Finite_Gij}. Thus, in general backgrounds it takes the form
	\begin{align}
		G_{\a\ad}{}^{i}{}_j =-{\frac\ri 4} \Big( [\cD_\a^{i},\bar{\cD}_{\ad j}] - \frac{1}{3} \d^i_j [\cD_\a^{k},\bar{\cD}_{\ad k}] \Big) \bold{\S} ~,
	\end{align}
	for some real scalar superfield $\bm{\S}$ inert under super-Weyl transformations. Hence, we conclude that $G_{\aa}{}^i{}_j$ is pure gauge.
	
	\subsection{$\sSU(3)$ superspace geometry}
	\label{Section5.3}
	
	As shown above, the torsion superfield $G_\aa{}^i{}_j$ describes purely gauge degrees of freedom.
	Thus, by employing the super-Weyl freedom described by eq. \eqref{6.17}, it may be gauged away
	\bea
	G_{\aa}{}^{i}{}_j=0~.
	\label{G2}
	\eea
	In this gauge, it is natural to shift $\cD_a$ as follows:
	\bea
	\label{6.19}
	\cD_a \longrightarrow \cD_a - \frac{\ri}{3} G_a {\mathbb Y}~.
	\eea
	Then, by making use of \eqref{U(2)algebra}, we find that these covariant derivatives obey the algebra:
	\begin{subequations} 
		\label{4.21}
		\bea
		\{ \cD_\a^i , \cD_\b^j \}
		&=&
		2 \ve_{\a \b} \ve^{ijk} \bar{W}_\ad \bar{\cD}^\ad_k
		+ 4 S^{ij}  M_{\a\b} 
		- 4 \ve_{\a \b} \ve^{ijk} Y^{\g \d}_k M_{\g \d}
		+ 2 \ve_{\a \b} \ve^{ijk} \bar{\cD}^\ad_k \bar{W}^\bd \bar{M}_{\ad \bd}
		\non \\
		&~&
		- 4\ve_{\a \b} S^{k[i} \mathbb{J}^{j]}{}_k
		+ 8 \ve^{kl(i}{Y}_{\a\b k}  \mathbb{J}^{j)}{}_l
		+ \ve_{\a \b} \ve^{ijk} \bar{\cD}^\ad_l \bar{W}_\ad \mathbb{J}^l{}_k
		\label{acr1} \\
		\{\cD_\a^i,\cDB^\bd_j\}&=&
		-2\ri\d^i_j \cD_\a{}^\bd
		+4\d^{i}_{j}G^{\d\bd}M_{\a\d}
		+4\d^{i}_{j}G_{\a\gd}\bar{M}^{\gd\bd}
		+8 G_\a{}^\bd \mathbb{J}^{i}{}_{j}~.~~~~~~~~~ \\
		\{ \cDB_i^\ad , \cDB_j^\bd \}
		&=&
		- 2 \ve^{\ad \bd} \ve_{ijk} W^\a \cD_\a^k
		- 4 \bar{S}_{ij}  \bar{M}^{\ad\bd} 
		+ 4 \ve^{\ad \bd} \ve_{ijk} \bar{Y}^{\gd \dd k} \bar{M}_{\gd \dd}
		+ 2 \ve^{\ad \bd} \ve_{ijk} {\cD}^{\a k} {W}^\b {M}_{\a \b}
		\non \\
		&~&
		+ 4\ve^{\ad \bd} \bar{S}_{k[i} \mathbb{J}^{k}{}_{j]}
		- 8 \ve_{kl(i} \bar{Y}^{\ad\bd k}  \mathbb{J}^{l}{}_{j)}
		- \ve^{\ad \bd} \ve_{ijk} \cD^{\a l} {W}_\ad \mathbb{J}^k{}_j
		\eea
	\end{subequations}
	
	The geometric superfields appearing above obey the Bianchi identities \eqref{BI-U2}  (upon imposing \eqref{G2}). Now, by examining equations \eqref{4.21}, we see that the $\sU(1)_R$ curvature has been eliminated and therefore the corresponding connection is flat. Consequently, it may be set to zero via an appropriate local $\sU(1)_R$ transformation; $\F_A =0$. As a result, the gauge group reduces to $\sSL( 2, {\mathbb C}) \times \sSU(3)_R$.
	
	It turns out that the gauge conditions \eqref{G2} and $\F_A=0$ allow for residual super-Weyl transformations, which are described
	by a parameter $\S$ constrained by
	\be
	\Big( [\cD_\a^{i},\bar{\cD}_{\ad j}] - \frac 1 3 \d^i_j [\cD_\a^{k},\bar{\cD}_{\ad k}] \Big) \S=0~.
	\label{Ucon}
	\ee
	The general solution of this condition is
	\bea
	\S = \frac{1}{2} (\s+\bar{\s})~, \qquad \bar{\cD}^\ad_i \s =0~,
	\qquad {\mathbb Y} \s =0~,
	\eea
	where the parameter $\s$ is covariantly chiral, with zero $\sU(1)_R$ charge, but otherwise arbitrary.
	To preserve the gauge condition $\F_A=0$, every super-Weyl transformation, eq. \eqref{6.17}, must be accompanied by the following compensating $\sU(1)_R$ transformation 
	\be
	\cD_A \longrightarrow \re^{-\frac{1}{12} (\s - \bar{\s}) \mathbb{Y}} \cD_A \re^{\frac{1}{12} (\s - \bar{\s}) \mathbb{Y}}~.
	\ee
	As a result, the algebra of covariant derivatives of $\sSU(3)$ superspace is preserved by the following set of super-Weyl transformations:
	\begin{subequations}
		\label{SU(N)sW}
		\begin{align}
			\cD_\a^{'i}&= \re^{\frac{1}{6} \s + \frac 1 3 \bar{\s}} \Big( \cD_\a^i+ \cD^{\b i}\s M_{\a \b} - \cD_{\a}^j\s \mathbb{J}^{i}{}_j \Big) ~, 
			\\ 
			\bar{\cD}_{i}^{' \ad}&=\re^{\frac{1}{3} \s + \frac{1}{6} \bar{\s}} \Big( \bar{\cD}^\ad_i-2 \bar{\cD}_{ \bd i} \bar{\s} \bar{M}^{\ad \bd} + \bar{\cD}^{\ad}_j \bar{\s} \mathbb{J}^{j}{}_i \Big)~,
			\\
			\cD_\aa' &= \re^{\hf \s + \hf \bar{\s}} \Big(\cD_\aa + \frac{\rm i}{2} \cD^i_{\a} \s \cDB_{\ad i} + \frac{\rm i}{2} \cDB_{\ad i} \bar{\s} \cD_{\a}^i + \hf \Big( \cD^\b{}_\ad (\s + \bar \s ) - \frac{\ri}{2} \cD^{\b i} \s \cDB_{\ad i} \bar{\s} \Big) M_{\a \b} \non \\ & + \hf \Big( \cD_{\a}{}^\bd (\s + \bar{\s}) + \frac{\ri}{2} \cD_{\a}^i \s \cDB^{\bd}_i \bar{\s} \Big) { \bar M}_{\ad \bd} 
			- \frac \ri 2 \cD_\a^i \s \cDB_{\ad j} \bar{\s} \mathbb{J}^j{}_i \Big) ~, \\
			W^{'}_\a&= \re^{\hf \s} W_\a
			\\
			S^{' ij}&= \re^{\frac{1}{3} \s + \frac 2 3 \bar{\s}} \Big( S^{ij}-{\frac14}\cD^{ij} \s + \frac 1 4 \cD^{\a (i} \s \cD_\a^{j)} \s \Big)~, 
			\label{super-Weyl-S} \\
			Y^{'}_{\a\b i}&= \re^{\frac{1}{3} \s + \frac 2 3 \bar{\s}} \Big( Y_{\a\b i}+{\frac18} \ve_{ijk} \cD_{(\a}^{j} \cD_{\b)}^{k}\s + \frac18 \ve_{ijk} \cD_{(\a}^{j} \s \cD_{\b)}^{k} \s \Big)~,
			\label{super-Weyl-Y} \\
			G_{\aa}' &=
			\re^{\frac12 \s + \frac12 \bar{\s}} \Big(
			G_{\a\bd} -{\frac{\ri}4}
			\cD_{\a \ad} (\s-\bar{\s})
			-\frac{1}{12} \cD_\a^i \s \bar{\cD}_{\ad i} \bar{\s}
			\Big)
			~.
			\label{super-Weyl-G}
		\end{align}
	\end{subequations}

\section{Discussion}
\label{Section6}

The primary outcome of this work is the construction of the four-dimensional $\cN=3$ conformal superspace, extending the $\cN=1$ and $\cN=2$ cases developed over ten years ago by Butter \cite{ButterN=1,ButterN=2}.
While this superspace geometry contains the full superconformal group as its structure group, it was shown in section \ref{Section5} that, by performing certain degauging procedures, the structure group may be reduced to $\sSL(2,\mathbb{C}) \times \sU(3)_R$, at which point one obtains the $\sU(3)$ superspace due to Howe \cite{Howe}. Actually, as shown in section \ref{Section5.3}, the structure group may be further reduced to $\sSL(2,\mathbb{C}) \times \sSU(3)_R$ by adopting a certain gauge.

By employing this manifestly superconformal framework, in section \ref{Section3.1} we described a superspace action principle based on a chiral Lagrangian $\cL_c$. As an application, we utilised this chiral action principle to propose a superspace action for conformal supergravity \eqref{3.9} and a supersymmetric extension of the Pontryagin term \eqref{3.11}.

Additionally, by an application of the ectoplasm method, in section \ref{Section4} we described an alternative action principle which is ideal for reducing a given chiral action to components. This was based on the use of a self-dual vector multiplet coupled to conformal supergravity, which was described in section \ref{Section4.2}.

While we have presented our results here within the realm of superspace, there exists an off-shell approach to $\cN=3$ conformal supergravity based on the superconformal tensor calculus, see \cite{vanMvanP,HMS}. In appendix \ref{AppendixB}, we provided the salient details of how the $\cN=3$ Weyl multiplet \cite{vanMvanP} manifests at the component level. It would be interesting to expand this analysis in the future.

Our conformal superspace setting opens up the possibility to formulate the dynamics of the off-shell ${\cal N}=3$ super Yang-Mills theory coupled to conformal supergravity. We remind the reader that the off-shell formulation for ${\cal N}=3$ super Yang-Mills theory in Minkowski superspace was developed by Galperin et al. within the $\cN=3$ harmonic superspace approach  \cite{GIKOS1, GIKOS2}. Technical details of coupling this theory to conformal supergravity will be discussed elsewhere. Here we restrict our discussion to showing how the multiplets introduced in this paper, including the super-Bach tensor $B^i{}_j$ and the Lagrangian $\mathscr{L}^{ij}$ defined by \eqref{4.8}, admit a natural interpretation in the framework of 
a curved analogue of the $\cN=3$  isotwistor superspace pioneered by Rosly 
\cite{Rosly} (see also \cite{RoslyS}). We extend the curved $\cN=3$ superspace $\cM^{4|12} $ with auxiliary bosonic dimensions and consider
\bea
\cM^{4|12} \times F (1,2, {\mathbb C}^3) ~,
\eea
where $F (1,2, {\mathbb C}^3) $ is a three-dimensional complex flag manifold. The latter is naturally realised as the quadric\footnote{The points of $ F (1,2, {\mathbb C}^3) $ are sequences $V_1 \subset V_2 \subset {\mathbb C}^3$, where $V_1$ and $V_2$ are one- and two-dimensional subspaces of ${\mathbb C}^3$. It is easy to see that the realisation \eqref{6.2} indeed describes the flag manifold.
In the case that $p_i$ and $q^i$ are given, the two-plane $V_2$, in which the one-dimensional 
subspace $V_1 = \left\{ \z p_i , ~ \z \in {\mathbb C} \right\} $ is embedded, is spanned by two three-vectors $p_i $ and $r_i$ such that $ \ve^{i j k} p_j r_k   \propto q^i$. }  
\bea
p_i q^i = 0  
\label{6.2}
\eea
in  ${\mathbb C}P^2 \times {\mathbb C}P^2$ parametrised by homogeneous coordinates $p_i$ and $q^i$, respectively. Associated with the spinor covariant derivatives $\nabla^i_\a$ and $\bar \nabla^\ad_i$ are the four fermionic operators
\bea
{\bm \nabla}_\a := p_i \nabla^i_\a~, \qquad \bar {\bm \nabla}^\ad := q^i \bar \nabla_i^\ad~.
\label{6.33}
\eea
In accordance with \eqref{algebra}, these operators anticommute with each other, 
\bea
\big\{ {\bm \nabla}_\a , {\bm \nabla}_\b \big\}=0~, \qquad \big\{ {\bm \nabla}_\a , \bar {\bm \nabla}_\bd \big\}=0~, \qquad 
\big\{ \bar {\bm \nabla}_\ad , \bar {\bm \nabla}_\bd \big\}=0~.
\eea
As a result, it is possible to introduce constrained primary supermultiplets  on $\cM^{4|12} \times F (1,2, {\mathbb C}^3) $ that are annihilated by the operators \eqref{6.33}, 
\bea
{\bm \nabla}_\a \F^{(m,n)} = 0 ~, \qquad \bar {\bm \nabla}_\ad \F^{(m,n)} =0~,
\eea
where $m$ and $n$ denote the degrees of homogeneity of $\F^{(m,n)}$  with respect to the variables $p_i$ and $q^i$, respectively, 
\bea
\F^{(m,n)} \Big(z, \,\z p , \,\z^{-1} \frac{\bar p}{p^\dagger p} , \,\x q, \,\x^{-1} \frac{\bar q }{qq^\dagger } \Big) 
= \z^m \x^n \F^{(m,n)} \Big(z, \, p , \, \frac{\bar p}{p^\dagger p} ,  \, q, \, \frac{\bar q }{qq^\dagger } \Big) ~.
\eea 
Here we have introduced the conjugate variables $\bar p^i := \overline{p_i}$ 
and $\bar q_i :=\overline{q^i}$.\footnote{Owing to the relations \eqref{A.5c} and \eqref{A.5d},
$\F^{(m,n)} $ does not carry any spinor indices.}

The supermultiplets of the above type,  which have appeared in this paper, are holomorphic in $p$'s and $q$'s, 
\bea
\F^{(m,n)} = \F^{(m,n)} \big(z,  p,    q \big) ~.
\label{6.7}
\eea
These are:
\begin{subequations}
\bea
\cW^{(1,0)} &=& \cW^i p_i ~, \qquad \qquad  \bar \cW^{(0,1)} = \bar \cW_i q^i~, \\
\mathscr{L}^{(2,0)} &=& \mathscr{L}^{ij} p_i p_j~, \qquad
\bar{\mathscr{L}}^{(0,2)} = \bar{\mathscr{L}}_{ij} q^i q^j ~, \\
B^{(1,1)} &=& B^i{}_j  p_i q^j~.
\eea
\end{subequations}
A holomorphic multiplet \eqref{6.7} is primary provided its dimension $\D$ and $\sU(1)$ charge $Q$ are as follows: 
\bea
\D =m+n~, \qquad Q = 2(n-m)~,
\eea
which follows from equations \eqref{A.5c} and \eqref{A.5d}.

The internal space $F (1,2, {\mathbb C}^3) $ can be realised as the coset space
\bea
F (1,2, {\mathbb C}^3) = \frac{\sSU(3)}{\sU(1) \times \sU(1)}~,
\eea
which is used within the $\cN=3$ harmonic superspace approach \cite{GIKOS1, GIKOS2}.
This coset space is parametrised in terms of harmonics ${\bm u}= (u_i{}^J) \equiv (u_i , v_i , w_i) \in \sSU(3)$ which are related to $p_i$ and $q^i$ as follows:
\bea
u_i:= \frac{p_i}{\sqrt{p^\dagger p}}~, \qquad v_i := \frac{\bar q_i }{\sqrt{qq^\dagger}} ~, \qquad 
\bar w^i = \ve^{ijk} u_j v_k~.
\eea
The homogeneous coordinates $p_i$ and $q^i$ for ${\mathbb C}P^2 \times {\mathbb C}P^2$ 
are defined modulo scale transformations $p_i \to \z p_i$ and $q^i \to \x q^i$, with $\z, \x \in {\mathbb C}\setminus 
\{ 0\}$. It follows that the harmonics $u_i$ and $v_i$ are defined modulo arbitrary phase transformations.

Recently, there has been much interest in $\cN=3$ superconformal field theories, which was initiated in 
Refs. \cite{Aharony:2015oyb, Garcia-Etxebarria:2015wns}. Since our paper has provided the $\cN=3$ conformal superspace setting, now it becomes possible to study the superconformal anomalies of such theories. 

Within the $\cN=3$ harmonic superspace approach, there have appeared interesting results on the $\cN=3$ 
supersymmetric Born-Infeld action \cite{Ivanov:2001ec} and scale/superconformal  invariant low-energy effective actions in $\cN = 3$ super Yang-Mills theory
\cite{Buchbinder:2004rj, Buchbinder:2011zu}. It would be interesting to couple these models to $\cN=3$ supergravity.

\noindent
{\bf Acknowledgements:} 
We are grateful to Gabriele Tartaglino-Mazzucchelli and Arkady Tseytlin for useful comments and suggestions. 
This work was supported in part by the Australian Research Council, project No. DP230101629.

\appendix

\section{The $\cN=3$ superconformal algebra}
\label{AppendixA}

In this appendix, we spell out our conventions for the superconformal algebra of $\cN = 3$ Minkowski superspace, $\mathfrak{su}(2,2|3)$. 

The conformal algebra, $\mathfrak{su}(2,2)$, consists of the translation $(P_a)$, Lorentz $(M_{ab})$, special conformal $(K_a)$ and dilatation $(\mathbb{D})$ generators. Amongst themselves, they obey the algebra
\begin{subequations} 
	\label{2.17}
	\begin{align}
		&[M_{ab},M_{cd}]=2\eta_{c[a}M_{b]d}-2\eta_{d[a}M_{b]c}~, \phantom{inserting blank space inserting} \\
		&[M_{ab},P_c]=2\eta_{c[a}P_{b]}~, \qquad \qquad \qquad \qquad ~ [\mathbb{D},P_a]=P_a~,\\
		&[M_{ab},K_c]=2\eta_{c[a}K_{b]}~, \qquad \qquad \qquad \qquad [\mathbb{D},K_a]=-K_a~,\\
		&[K_a,P_b]=2\eta_{ab}\mathbb{D}+2M_{ab}~.
	\end{align}
\end{subequations}

The $R$-symmetry group $\sU(3)_R$ is generated by the $\sU(1)_R$ $(\mathbb{Y})$ and $\sSU(3)_R$ $(\mathbb{J}^i{}_j)$ generators, which commute with all elements of the conformal algebra. Amongst themselves, they obey the commutation relations
\begin{align}
	[\mathbb{J}^{i}{}_j,\mathbb{J}^{k}{}_l] = \d^i_l \mathbb{J}^k{}_j - \d^k_j \mathbb{J}^i{}_l ~.
\end{align}

The superconformal algebra is then obtained by extending the translation generator to $P_A=(P_a,Q_\a^i,\bar{Q}^\ad_i)$ and the special conformal generator to $K^A=(K^a,S^\a_i,\bar{S}_\ad^i)$. The commutation relations involving the $Q$-supersymmetry generators with the bosonic ones are:
\begin{subequations} 
	\bea
	\big[M_{ab}, Q_\g^i \big] &=& (\s_{ab})_\g{}^\d Q_\d^i ~,\quad 
	\big[M_{ab}, \bar Q^\gd_i \big] = (\tilde{\s}_{ab})^\gd{}_\dd \bar Q^\dd_i~,\\
	\big[\mathbb{D}, Q_\a^i \big] &=& \hf Q_\a^i ~, \quad
	\big[\mathbb{D}, \bar Q^\ad_i \big] = \hf \bar Q^\ad_i ~, \\
	\big[\mathbb{Y}, Q_\a^i \big] &=&  Q_\a^i ~, \quad
	\big[\mathbb{Y}, \bar Q^\ad_i \big] = - \bar Q^\ad_i ~, \label{2.19c} \\
	\big[\mathbb{J}^i{}_j, Q_\a^k \big] &=&  - \d^k_j Q_\a^i + \frac{1}{3} \d^i_j Q_\a^k ~, \quad
	\big[\mathbb{J}^i{}_j, \bar Q^\ad_k \big] = \d^i_k \bar Q^\ad_j - \frac{1}{3} \d^i_j \bar Q^\ad_k ~,  \\
	\big[K^a, Q_\b^i \big] &=& -\ri (\s^a)_\b{}^\bd \bar{S}_\bd^i ~, \quad 
	\big[K^a, \bar{Q}^\bd_i \big] = 
	-\ri ({\s}^a)^\bd{}_\b S^\b_i ~.
	\eea
\end{subequations}
At the same time, the commutation relations involving the $S$-supersymmetry generators 
with the bosonic operators are: 
\begin{subequations}
	\bea
	\big [M_{ab} , S^\g_i \big] &=& - (\s_{ab})_\b{}^\g S^\b_i ~, \quad
	\big[M_{ab} , \bar S_\gd^i \big] = - (\ts_{ab})^\bd{}_\gd \bar S_\bd^i~, \\
	\big[\mathbb{D}, S^\a_i \big] &=& -\hf S^\a_i ~, \quad
	\big[\mathbb{D}, \bar S_\ad^i \big] = -\hf \bar S_\ad^i ~, \\
	\big[\mathbb{Y}, S^\a_i \big] &=&  -S^\a_i ~, \quad
	\big[\mathbb{Y}, \bar S_\ad^i \big] =  \bar S_\ad^i ~,  \label{2.20c}\\
	\big[\mathbb{J}^i{}_j, S^\a_k \big] &=&  \d^i_k S^\a_j - \frac{1}{3} \d^i_j S^\a_k ~, \quad
	\big[\mathbb{J}^i{}_j, \bar S_\ad^k \big] = - \d_j^k \bar S_\ad^i + \frac{1}{3} \d^i_j \bar S_\ad^k ~,  \\
	\big[ S^\a_i , P_b \big] &=& \ri (\s_b)^\a{}_\bd \bar{Q}^\bd_i ~, \quad 
	\big[\bar{S}_\ad^i , P_b \big] = 
	\ri ({\s}_b)_\ad{}^\b Q_\b^i ~.
	\eea
\end{subequations}
Finally, the anti-commutation relations of the fermionic generators are: 
\begin{subequations}
	\label{A.5}
	\bea
	\{Q_\a^i , \bar{Q}^\ad_j \} &=& - 2 \ri \d^i_j (\s^b)_\a{}^\ad P_b=- 2 \ri \d^i_j  P_\a{}^\ad~, \\
	\{ S^\a_i , \bar{S}_\ad^j \} &=& 2 \ri  \d_i^j (\s^b)^\a{}_\ad K_b=2 \ri \d_i^j  K^\a{}_\ad
	~, \\
	\{ S^\a_i , Q_\b^j \} &=& \d_i^j \d^\a_\b \Big(2 \mathbb{D} - \frac{1}{3}\mathbb{Y} \Big) - 4 \d_i^j  M^\a{}_\b 
	+ 4 \d^\a_\b  \mathbb{J}^j{}_i ~, \label{A.5c}\\
	\{ \bar{S}_\ad^i , \bar{Q}^\bd_j \} &=& \d_j^i \d^\bd_\ad \Big(2 \mathbb{D} + \frac{1}{3} \mathbb{Y} \Big) + 4 \d_j^i  \bar{M}_\ad{}^\bd 
	- 4 \d_\ad^\bd  \mathbb{J}^i{}_j  ~. \label{A.5d}
	\eea
\end{subequations}
We emphasise that all (anti-)commutators not listed above vanish identically. 

It is convenient to group the superconformal generators introduced above into two disjoint subsets; $P_A$ and the non-translational generators $X_{\underline A}= (M_{ab}, \mathbb{J}^i{}_j, \mathbb{Y}, \mathbb{D}, K^A)$. Then, the superconformal algebra relations above can be succinctly written as
\begin{align}
	\label{StructureConstants}
	\big[ X_{\underline A} , X_{\underline B} \big \} = - f_{{\underline A} {\underline B}}{}^{\underline C} X_{\underline C} ~, \qquad \big[ X_{\underline A} , P_{B} \big \} = - f_{{\underline A} {B}}{}^{\underline C} X_{\underline C} - f_{{\underline A} {B}}{}^{C} P_{C}~,
\end{align}
where $f_{{\underline A} {\underline B}}{}^{\underline C}$, $f_{{\underline A} {B}}{}^{\underline C}$, and $f_{{\underline A} {B}}{}^{C}$ are the structure constants of the superconformal algebra and may be read off from equations \eqref{2.17} - \eqref{A.5}.

\section{The $\cN=3$ Weyl multiplet via superspace}
\label{AppendixB}

In section \ref{Section2} we have computed the geometry of a superspace with the $\cN=3$ superconformal group as its local structure group. 
As it may be utilised to describe conformal supergravity, it is instructive to verify that it encodes the Weyl multiplet described in \cite{vanMvanP}. The latter involves a set of gauge one-forms, namely: the vielbein $e_m{}^a$, gravitini $\psi_m{}^\a_i$, $\sU(1)_R$ gauge field $\cA_m$, $\sSU(3)_R$ gauge field $\cV_m{}^i{}_j$, and dilatation gauge field $b_m$. Modulo purely gauge degrees of freedom, they may be shown to be the only independent geometric fields and arise as the lowest dimensional components of the following superforms
\begin{subequations}
\label{B.1}
\begin{align}
	e_m{}^a &:= E_m{}^a| ~, \qquad \psi_m{}^\a_i := 2E_m{}^\a_i| ~, \\
	\cA_m &:= \F_m| ~, \qquad \cV_m{}^i{}_j := \F_m{}^i{}_j| ~, \qquad b_m := B_m|~,
\end{align}
\end{subequations}
where we have made use of the shorthand $U| := U|_{\q = \bar{\q} = 0}$ to denote the bar projection of a covariant $\cN=3$ superfield $U$.\footnote{This should not be confused with the $\cN=2$ projection introduced at the end of section \ref{section4.3}.}
For later convenience, we also define the following composite fields
\begin{align}
	\label{B.2}
	\o_m{}^{bc} := \O_{m}{}^{bc} |_{\q = \bar{\q} = 0} ~, \qquad \vf_m{}^i_\a := \mathfrak{F}_m{}^i_\a |_{\q = \bar{\q} = 0}~.
\end{align}

As discussed in \cite{vanMvanP}, to ensure that the local superconformal transformations of the Weyl multiplet close off-shell, it is necessary to introduce a set of covariant `matter' fields. These appear as component fields of the super-Weyl tensor $W_\a$:
\begin{subequations}
	\label{ComponentMatterFields}
	\begin{align}
		\L_\a &:= W_\a | ~, \qquad K^a \L_\a = 0 ~, \qquad \mathbb{D} \L_\a = \frac{1}{2} \L_\a~, \qquad \mathbb{Y} \L_\a = -3 \L_\a ~, \\
		E^i &:= \nabla^{\a i} W_\a | ~, \qquad K^a E^i = 0 ~, \qquad \mathbb{D} E^i = E^i~, \qquad \mathbb{Y} E^i = -2 E^i ~, \\
		T^i_{\a \b} &:= \nabla^{i}_{(\a} W_{\b)} | ~, \qquad K^a T^i_{\a \b} = 0 ~, \qquad \mathbb{D} T^i_{\a \b} = T^i_{\a \b}~, \qquad \mathbb{Y} T^i_{\a \b} = -2 T^i_{\a \b} ~, \\
		\chi^{ij}_{\a} &:= \nabla^{ij} W_{\a} | ~, \qquad K^a \chi^{ij}_{\a} = 0 ~, \qquad \mathbb{D} \chi^{ij}_{\a} = \frac{3}{2} \chi^{ij}_{\a}~, \qquad \mathbb{Y} \chi^{ij}_{\a} = - \chi^{ij}_{\a} ~, \\
		\z^{\a}_{i} &:= \ve_{ijk} \nabla^{\a j}  \nabla^{\b k} W_{\b} | ~, \qquad K^a \z^{\a}_{i} = 0 ~, \qquad \mathbb{D} \z^{\a}_{i} = \frac{3}{2} \z^{\a}_{i}~, \qquad \mathbb{Y} \z^{\a}_{i} = - \z^{\a}_{i} ~, \\
		D^{i}{}_{j} &:= B^i{}_j | ~, \qquad K^a D^{i}{}_{j} = 0 ~, \qquad \mathbb{D} D^{i}{}_{j} = 2 D^{i}{}_{j} ~, \qquad \mathbb{Y} D^{i}{}_{j} = 0 ~.
	\end{align}
\end{subequations}
Inspecting the above equations, we see that not all components of $W_\a$ are accounted for. This is because, similar to the $\cN=1$ and $\cN=2$ cases, the remaining component fields may be shown to be functions of the connections \eqref{B.1}. This will be described elsewhere.

To conclude, we now spell out the $Q$-supersymmetry transformation laws for the Weyl multiplet \eqref{B.1}. We recall that such a transformation, parametrised by the spinor $\e^\a_i$ and its conjugate $\bar{\e}_\ad^i$, should be identified with a supergravity gauge transformation \eqref{66.2} with
\begin{align}
	\label{B.3}
	\mathscr{K}(\e) = \xi^\a_i \nabla_\a^i + \bar{\xi}_\ad^i \bar{\nabla}^\ad_i ~, \qquad \qquad \e^\a_i = \xi^\a_i|_{\q = \bar{\q} = 0} ~.
\end{align}
Making use of eq. \eqref{connectionTfs}, a routine calculation leads to the following transformation laws:\footnote{While we do not spell out the transformation laws for the covariant matter fields \eqref{ComponentMatterFields}, it should be emphasised that they may be readily computed via equation \eqref{6.3}, in conjunction with \eqref{B.3}.}
\begin{subequations}
	\label{QSusy}
	\begin{align}
		\d_{\e} e_m{}^a &= \ri \Big( \e_i \s^a \bar{\psi}_m^i - \psi_{m i} \s^a \bar{\e}^i \Big) ~, \\
		\d_{\e} \psi_m{}^\a_i &= \Big(\partial_m - \o_m{}^{bc} M_{bc} + \ri \cA_m - \cV_m{}^k{}_j \mathbb{J}^j{}_k + \frac 1 2 b_m \Big) \e^\a_i \non \\
		&\phantom{=}~- \frac{\ri}{2} \ve_{ijk} \big( T^\a{}_\b{}^j + \hf \d^\a_\b E^j \big) (\s_m \bar{\e}^k)^\b - (\e_i \s_m \bar{\L}) \L^\a~, \\
		\d_{\e} \cA_m &= - \frac \ri 3 \varphi_{m}{}^i_\a \e^\a_i - \frac \ri {12} \ve^{ijk} \e^\a_i \psi_{m \a j} \bar{E}_k + \frac{1}{24} \e_i \s_m \bar{\z}^i - \frac{1}{12} (\e_i \s_m \bar{\L}) E^i + \text{c.c.} ~, \\
		\d_{\e} \cV_m{}^i{}_j &= -4 \vf_m{}^i_\a \e^\a_j + \ve^{kli} \e^\a_k \psi_{m \a l} \bar{E}_j + \frac{\ri}{4} \ve^{kli} (\e_k \s_m \bar{\chi}_{lj}) - \frac{\ri}{4} \e_j \s_m \bar{\z}^i - \ri (\e_j \s_m \bar{\L}) E^i \non \\
		&\phantom{=}~ - \text{trace} - \text{h.c.} ~, \\
		\d_{\e} b_m &= - 2 \vf_{m}{}^i_\a \e^\a_i - \ve^{ijk} \e^\a_i \psi_{m \a j} \bar{E}_k - \frac{\ri}{4} (\e_i \s_m \bar{\z}^i) - \frac{\ri}{2} (\e_i \s_m \bar{\L}) E^i + \text{c.c.}
	\end{align}
\end{subequations}
We note that these transformations differ slightly from those presented in \cite{vanMvanP}, see also \cite{HS}, however, as will be shown below, these two sets of transformations are equivalent.

To prove this statement, it is first necessary to shift the composite fields \eqref{B.2} as follows:
\begin{align}
	\o_m{}^{\a}{}_\b \quad &\longrightarrow \quad \o_m{}^{\a}{}_\b + 2 (\s_m \bar{\L})^\a \L_\b - \d^\a_\b (\s_m \bar{\L})^\g \L_\g ~,\\
	\vf_m{}^\a_i \quad &\longrightarrow \quad \vf_m{}^\a_i + \ri \bar{\L}_\bd  (\tilde{\s}_m)^{\bd \b} T_{\b \a}^i+ \frac 1 4 \ve^{ijk} \psi_{m \a j} \bar{E}_k ~.
\end{align}
Then, supplementing each $Q$-supersymmetry transformation \eqref{B.3} with the following $\e$-dependent special superconformal transformation
\begin{subequations}
\begin{align}
	& \qquad \qquad \mathscr{K}(\e) \quad \longrightarrow \quad \mathscr{K}(\e) + \eta_B K^B ~, \\ 
	\qquad \eta_\a^i | &= - \ri \L_\a \bar{\e}_\ad^i \bar{\L}^\ad - \frac 1 4 \ve^{ijk} \e_{\a j} \bar{E}_k ~, \\
	\eta_a | &= - \frac 3 8 \ve^{ijk} \e^\a_i \psi_{a \a j} \bar{E}_k - \frac \ri 8 \e_i \s_a \bar{\z}^i - \frac \ri 4 (\e_i \s_a \bar{\L}) E^i + \frac \ri 2 \psi_{a}{}^\a_i \L_\a \bar{\e}_\ad^i \bar{\L}^\ad \non \\
	&\phantom{=}~ - \ri \bar{\L}_\bd (\tilde{\s}_a)^{\bd \b} T_{\b \a}^i \e^\a_i + \text{c.c.} ~,
\end{align}
\end{subequations}
leads to `improved' transformations for the Weyl multiplet:
\begin{subequations}
	\label{QSusy2}
	\begin{align}
		\d_{\e} e_m{}^a &= \ri \Big( \e_i \s^a \bar{\psi}_m^i - \psi_{m i} \s^a \bar{\e}^i \Big) ~, \\
		\d_{\e} \psi_m{}^\a_i &= \Big(\partial_m - \o_m{}^{bc} M_{bc} + \ri \cA_m - \cV_m{}^k{}_j \mathbb{J}^j{}_k + \frac 1 2 b_m \Big) \e^\a_i \non \\
		&\phantom{=}~- \frac{\ri}{2} \ve_{ijk}T^\a{}_\b{}^j (\s_m \bar{\e}^k)^\b + \ve_{ijk} \bar{\e}_\ad^j \bar{\psi}^{\ad k}_m \L^\a ~, \\
		\d_{\e} \cA_m &= - \frac \ri 3 \varphi_{m}{}^i_\a \e^\a_i - \frac \ri {24} \ve^{ijk} \e^\a_i \psi_{m \a j} \bar{E}_k + \frac{1}{24} \e_i \s_m \bar{\z}^i + \frac{1}{12} (\e_i \s_m \bar{\L}) E^i + \text{c.c.} ~, \\
		\d_{\e} \cV_m{}^i{}_j &= -4 \vf_m{}^i_\a \e^\a_j - \hf \ve^{kli} \e^\a_k \psi_{m \a l} \bar{E}_j + \frac{\ri}{4} \ve^{kli} (\e_k \s_m \bar{\chi}_{lj}) - \frac{\ri}{4} \e_j \s_m \bar{\z}^i - \ri (\e_j \s_m \bar{\L}) E^i \non \\
		&\phantom{=}~ - \frac{16 \ri}{3} \bar{\L}_\ad (\tilde{\s}_m)^{\ad \a} T_{\a \b}^i \e^\b_j + 2 \ri \psi_m{}^\a_j \L_\a \bar{\e}_\ad^i \bar{\L}^\ad - \text{trace} - \text{h.c.} ~, \\
		\d_{\e} b_m &= - 2 \vf_{m}{}^i_\a \e^\a_i + \text{c.c.}
	\end{align}
\end{subequations}
These transformation laws are in agreement with those of \cite{vanMvanP,HS}, which proves our claim.

\begin{footnotesize}

\end{footnotesize}


\begin{thebibliography}{66}



\bibitem{KTvN} 
M.~Kaku, P.~K.~Townsend and P.~van Nieuwenhuizen,
``Properties of conformal supergravity,''
Phys.\ Rev.\ D {\bf 17}, 3179 (1978);
P.~K.~Townsend and P.~van Nieuwenhuizen,
``Simplifications of conformal cupergravity,''
Phys. Rev. D \textbf{19}, 3166 (1979).

\bibitem{BdRdW}
E.~Bergshoeff, M.~de Roo and B.~de Wit,
``Extended conformal supergravity,''
Nucl. Phys. B \textbf{182}, 173 (1981).


\bibitem{Butter:2016mtk}
D.~Butter, F.~Ciceri, B.~de Wit and B.~Sahoo,
``Construction of all N=4 conformal supergravities,''
Phys. Rev. Lett. \textbf{118}, no.8, 081602 (2017)
[arXiv:1609.09083 [hep-th]].

\bibitem{Butter:2019edc}
D.~Butter, F.~Ciceri and B.~Sahoo,
``$N=4$ conformal supergravity: the complete actions,''
JHEP \textbf{01}, 029 (2020)
[arXiv:1910.11874 [hep-th]].

  
\bibitem{KakuTownsend}
M.~Kaku, P.~K.~Townsend,
``Poincar\'e supergravity as broken superconformal gravity,''
Phys.\ Lett.\  {\bf B76}, 54 (1978).

 
\bibitem{FVP} 
D.~Z.~Freedman and A.~Van Proeyen,
 {\it Supergravity}, Cambridge, UK: Cambridge Univ. Press (2012) 607 p.  

\bibitem{Ferrara:1977ij}
S.~Ferrara, M.~Kaku, P.~K.~Townsend and P.~van Nieuwenhuizen,
``Gauging the graded conformal group with unitary internal symmetries,''
Nucl. Phys. B \textbf{129}, 125 (1977).

\bibitem{deWit:1978pd}
B.~de Wit and S.~Ferrara,
``On higher order invariants in extended supergravity,''
Phys. Lett. B \textbf{81}, 317 (1979).
  
 \bibitem{deRoo}
M.~de Roo,
``Matter coupling in N=4 supergravity,''
Nucl. Phys. B \textbf{255}, 515 (1985);
``Gauged N=4 matter couplings,''
Phys. Lett. B \textbf{156}, 331(1985).
  
\bibitem{FT84}
E.~S.~Fradkin and A.~A.~Tseytlin,
``Conformal anomaly in Weyl theory and anomaly free superconformal theories''
Phys. Lett. B \textbf{134}, 187 (1984);
``Instanton zero modes and $\beta$-functions in conformal supergravity,''
Phys. Lett. B \textbf{134}, 307 (1984).

\bibitem{FT85} 
E.~S.~Fradkin and A.~A.~Tseytlin,
  ``Conformal supergravity,''
  Phys.\ Rept.\  {\bf 119}, 233 (1985).

\bibitem{Romer:1985yg}
H.~R\"omer and P.~van Nieuwenhuizen,
``Axial anomalies in $N=4$ conformal supergravity,''
Phys. Lett. B \textbf{162}, 290 (1985).

\bibitem{Tseytlin}
A.~A.~Tseytlin,
``On divergences in non-minimal $N=4$ conformal supergravity,''
J. Phys. A \textbf{50}, no.48, 48LT01 (2017)
[arXiv:1708.08727 [hep-th]].  


 \bibitem{ZuminoSS} 
B.~Zumino,
  ``Supergravity and superspace,''
in {\it Recent Developments in  Gravitation - Carg\`ese 1978}, 
M. L\'evy and S. Deser (Eds.), N.Y., Plenum Press, 1979, pp. 405.

 
\bibitem{OS}
  V.~Ogievetsky and E.~Sokatchev,
  ``Structure of supergravity group,''
  Phys.\ Lett.\ B {\bf 79}, 222 (1978).

 \bibitem{HT}
P.~S.~Howe and R.~W.~Tucker,
``Scale invariance in superspace,''
Phys.\ Lett.\ B {\bf 80}, 138 (1978).


\bibitem{Siegel78}
W.~Siegel,
``Solution to constraints in Wess-Zumino supergravity formalism,''
Nucl.\ Phys.\  B {\bf 142}, 301 (1978). 

   	
\bibitem{ButterN=1}
D.~Butter, ``N=1 Conformal Superspace in Four Dimensions,''
Annals Phys. \textbf{325}, 1026 (2010)
[arXiv:0906.4399 [hep-th]].
		
\bibitem{ButterN=2}
D.~Butter, ``N=2 Conformal Superspace in Four Dimensions,''
JHEP \textbf{10}, 030 (2011)
[arXiv:1103.5914 [hep-th]].

\bibitem{Review1}
S.~M.~Kuzenko, E.~S.~N.~Raptakis and G.~Tartaglino-Mazzucchelli,
``Superspace approaches to $\mathcal{N}=1$ supergravity,''
[arXiv:2210.17088 [hep-th]].

\bibitem{Review2}
S.~M.~Kuzenko, E.~S.~N.~Raptakis and G.~Tartaglino-Mazzucchelli,
``Covariant superspace approaches to ${\cal N}=2$ supergravity,''
[arXiv:2211.11162 [hep-th]].

\bibitem{GIKOS}
A.~S.~Galperin, E.~A.~Ivanov, S.~N.~Kalitzin, V.~Ogievetsky, E.~Sokatchev, 
``Unconstrained N=2 matter, Yang-Mills and supergravity theories in harmonic
superspace,'' Class.\ Quant.\ Grav.\  {\bf 1}, 469 (1984).


\bibitem{GIOS}
  A.~S.~Galperin, E.~A.~Ivanov, V.~I.~Ogievetsky and E.~S.~Sokatchev,
{\it Harmonic Superspace},
Cambridge University Press,  Cambridge, 2001.





\bibitem{LR1}
U.~Lindstr\"om and M.~Ro\v{c}ek,
``New hyperk\"ahler  metrics  and new supermultiplets,''
  Commun.\ Math.\ Phys.\  {\bf 115}, 21 (1988).

\bibitem{LR2}
U.~Lindstr\"om and M.~Ro\v{c}ek,  
 ``N=2 super Yang-Mills theory in projective superspace,''
Commun.\ Math.\ Phys.\  {\bf 128}, 191 (1990).


\bibitem{GIKOS1}
A.~Galperin, E.~Ivanov, S.~Kalitzin, V.~Ogievetsky and E.~Sokatchev,
``Unconstrained off-shell N=3 supersymmetric Yang-Mills theory,''
Class. Quant. Grav. \textbf{2}, 155 (1985).

\bibitem{GIKOS2}
A.~Galperin, E.~Ivanov, S.~Kalitzin, V.~Ogievetsky and E.~Sokatchev,
``N=3 supersymmetric gauge theory,''
Phys. Lett. B \textbf{151}, 215 (1985).

\bibitem{RoslyS} A.~A.~Rosly and A.~S.~Schwarz,
  ``Supersymmetry in a space with auxiliary dimensions,''
 Commun.\ Math.\ Phys.\  {\bf 105}, 645 (1986).

 \bibitem{Galperin:1986id}
A.~S.~Galperin, E.~A.~Ivanov and V.~I.~Ogievetsky,
``Superspaces for $N=3$ supersymmetry,''
Sov. J. Nucl. Phys. \textbf{46} (1987), 543 [Yad.Fiz. 46 (1987) 948].

 \bibitem{vanMvanP}
J.~van Muiden and A.~Van Proeyen,
``The $ \mathcal{N} $ = 3 Weyl multiplet in four dimensions,''
JHEP \textbf{01}, 167 (2019)
[arXiv:1702.06442 [hep-th]].


\bibitem{HS}
S.~Hegde and B.~Sahoo,
``Comment on \textquotedblleft{}The $N$=3 Weyl multiplet in four dimensions\textquotedblright{},''
Phys. Lett. B \textbf{791}, 92 (2019)
[arXiv:1810.05089 [hep-th]].


\bibitem{HMS}
S.~Hegde, M.~Mishra and B.~Sahoo,
``N = 3 conformal supergravity in four dimensions,''
JHEP \textbf{04}, 001 (2022) [arXiv:2104.07453 [hep-th]].



\bibitem{BKNT-M1}
D.~Butter, S.~M.~Kuzenko, J.~Novak and G.~Tartaglino-Mazzucchelli,
``Conformal supergravity in three dimensions: New off-shell formulation,''
JHEP \textbf{09}, 072 (2013)
[arXiv:1305.3132 [hep-th]].

\bibitem{BKNT-M2}
D.~Butter, S.~M.~Kuzenko, J.~Novak and G.~Tartaglino-Mazzucchelli,
``Conformal supergravity in three dimensions: Off-shell actions,''
JHEP \textbf{10}, 073 (2013)
[arXiv:1306.1205 [hep-th]].


\bibitem{KNT-M}
S.~M.~Kuzenko, J.~Novak and G.~Tartaglino-Mazzucchelli,
``N=6 superconformal gravity in three dimensions from superspace,''
JHEP \textbf{01}, 121 (2014)
[arXiv:1308.5552 [hep-th]].

\bibitem{BKNT-M15}
  D.~Butter, S.~M.~Kuzenko, J.~Novak and G.~Tartaglino-Mazzucchelli,
``Conformal supergravity in five dimensions: New approach and applications,''
JHEP {\bf 1502}, 111 (2015).
[arXiv:1410.8682 [hep-th]].

\bibitem{BKNT}
  D.~Butter, S.~M.~Kuzenko, J.~Novak and S.~Theisen,
  ``Invariants for minimal conformal supergravity in six dimensions,''
  JHEP {\bf 1612}, 072 (2016)
  [arXiv:1606.02921 [hep-th]].


\bibitem{Howe}
P.~S.~Howe,
``A superspace approach to extended conformal supergravity,''
Phys.\ Lett.\ B {\bf 100}, 389 (1981);
``Supergravity in superspace,''  Nucl.\ Phys.\  B {\bf 199}, 309 (1982).

 \bibitem{Muller} M. M\"uller, {\it Consistent Classical Supergravity Theories},
(Lecture Notes in Physics, Vol. 336),
Springer, Berlin, 1989. 


\bibitem{HL20}
P.~S.~Howe and U.~Lindstr\"om,
``Superconformal geometries and local twistors,''
JHEP \textbf{04}, 140 (2021) [arXiv:2012.03282 [hep-th]].

 \bibitem{Sohnius:1978wk}
M.~F.~Sohnius,
``Bianchi identities for supersymmetric gauge theories,''
Nucl. Phys. B \textbf{136}, 461 (1978).

\bibitem{HST}
P.~S.~Howe, K.~S.~Stelle and P.~K.~Townsend, ``Supercurrents,''
Nucl. Phys. B \textbf{192}, 332 (1981).

\bibitem{HKR}
D.~Hutchings, S.~M.~Kuzenko and E.~S.~N.~Raptakis,
``The N=2 superconformal gravitino multiplet,''
Phys. Lett. B \textbf{845}, 138132 (2023)
[arXiv:2305.16029 [hep-th]].


 \bibitem{Siegel}
W.~Siegel,
``On-shell O($N$) supergravity in superspace,''
Nucl. Phys. B \textbf{177}, 325 (1981).

\bibitem{Kuzenko:2023ebe}
S.~M.~Kuzenko and E.~S.~N.~Raptakis,
``Self-duality for N-extended superconformal gauge multiplets,''
Nucl. Phys. B \textbf{997}, 116378 (2023)
[arXiv:2308.10660 [hep-th]].
 
\bibitem{Kuzenko:2021qcx}
S.~M.~Kuzenko and E.~S.~N.~Raptakis,
``Duality-invariant superconformal higher-spin models,''
Phys. Rev. D \textbf{104}, no.12, 125003 (2021)
[arXiv:2107.02001 [hep-th]].
  
  





  

\bibitem{Hasler}
M.~F.~Hasler, ``The three form multiplet in N=2 superspace,''
  Eur.\ Phys.\ J.\ C {\bf 1}, 729 (1998) [hep-th/9606076].


\bibitem{Gates}
S.~J.~Gates Jr., ``Ectoplasm has no topology: The prelude,''
in {\it Supersymmetries and Quantum Symmetries},
J. Wess and E. A. Ivanov (Eds.), Springer, Berlin, 1999, p. 46, arXiv:hep-th/9709104;
``Ectoplasm has no topology,''
Nucl.\ Phys.\  B {\bf 541}, 615 (1999)
[arXiv:hep-th/9809056].

\bibitem{GGKS}
S.~J.~Gates Jr., M.~T.~Grisaru, M.~E.~Knutt-Wehlau and W.~Siegel,
``Component actions from curved superspace: Normal coordinates and
ectoplasm,'' Phys.\ Lett.\  B {\bf 421}, 203 (1998)
[hep-th/9711151].


\bibitem{Castellani} 
  L.~Castellani, R.~D'Auria and P.~Fre,
{\it Supergravity and superstrings: A Geometric perspective. Vol. 2: Supergravity},
World Scientific,  Singapore, 1991, pp. 680--684. 

\bibitem{Arias:2014ona}
C.~Arias, W.~D.~Linch, III and A.~K.~Ridgway,
``Superforms in six-dimensional superspace,''
JHEP \textbf{05}, 016 (2016)
[arXiv:1402.4823 [hep-th]].

\bibitem{Gates:2014cqa}
S.~J.~Gates, W.~D.~Linch and S.~Randall,
``Superforms in five-dimensional $N = 1$ superspace,''
JHEP \textbf{05}, 049 (2015)
[arXiv:1412.4086 [hep-th]].

\bibitem{Linch:2014iza}
W.~D.~Linch and S.~Randall,
``Superspace de Rham complex and relative cohomology,''
JHEP \textbf{09}, 190 (2015)
[arXiv:1412.4686 [hep-th]].

\bibitem{GKT-M}
S.~J.~Gates, Jr., S.~M.~Kuzenko and G.~Tartaglino-Mazzucchelli,
``Chiral supergravity actions and superforms,''
Phys. Rev. D \textbf{80}, 125015 (2009)
[arXiv:0909.3918 [hep-th]].


\bibitem{KKR}
N.~E.~Koning, S.~M.~Kuzenko and E.~S.~N.~Raptakis,
``Embedding formalism for $ \mathcal{N} $-extended AdS superspace in four dimensions,''
JHEP \textbf{11}, 063 (2023)
[arXiv:2308.04135 [hep-th]].



\bibitem{Rosly}
 A.~A.~Rosly, ``Super Yang-Mills  constraints 
as integrability conditions,'' in {\it Proceedings of the International 
Seminar on Group Theoretical 
Methods in Physics},'' (Zvenigorod, USSR, 1982),
M. A. Markov  (Ed.), 
Nauka, Moscow, 1983, Vol. 1, p. 263 (in Russian);
English translation: in {\it Group Theoretical 
Methods in Physics},'' M. A. Markov, V. I. Man'ko 
and A. E. Shabad  (Eds.), Harwood Academic Publishers, 
London, Vol. 3, 1987, p. 587.

\bibitem{Aharony:2015oyb}
O.~Aharony and M.~Evtikhiev,
``On four dimensional N = 3 superconformal theories,''
JHEP \textbf{04}, 040 (2016)
[arXiv:1512.03524 [hep-th]].


\bibitem{Garcia-Etxebarria:2015wns}
I.~Garcia-Etxebarria and D.~Regalado,
``$ \mathcal{N}=3 $ four dimensional field theories,''
JHEP \textbf{03}, 083 (2016)
[arXiv:1512.06434 [hep-th]].

\bibitem{Ivanov:2001ec}
E.~A.~Ivanov and B.~M.~Zupnik,
``N=3 supersymmetric Born-Infeld theory,''
Nucl. Phys. B \textbf{618}, 3-20 (2001)
[arXiv:hep-th/0110074 [hep-th]].
		
\bibitem{Buchbinder:2004rj}
I.~L.~Buchbinder, E.~A.~Ivanov, I.~B.~Samsonov and B.~M.~Zupnik,
``Scale invariant low-energy effective action in N = 3 SYM theory,''
Nucl. Phys. B \textbf{689}, 91-107 (2004)
[arXiv:hep-th/0403053 [hep-th]].

\bibitem{Buchbinder:2011zu}
I.~L.~Buchbinder, E.~A.~Ivanov, I.~B.~Samsonov and B.~M.~Zupnik,
``Superconformal N=3 SYM Low-Energy Effective Action,''
JHEP \textbf{01}, 001 (2012)
[arXiv:1111.4145 [hep-th]].
		
\end{thebibliography}
\end{document}